\documentclass{emulateapj-rtx4}

\newcommand{\sdss}{{\small {SDSS}}}

\newcommand{\sdssii}{{\small {SDSS-II}}}

\newcommand{\snls}{{\small {SNLS}}}
\newcommand{\essence}{{\small {ESSENCE}}}
\newcommand{\sn}{{\small {SN}}}
\newcommand{\ccsn}{{\small {CC~SN}}}

\newcommand{\snia}{{\small {SN~I}}a}
\newcommand{\snibc}{{\small {SN~I}}b/c}
\newcommand{\snii}{{\small {SN~II}}}

\newcommand{\photoz}{photo-$z$}
\newcommand{\suspect}{{\small {SUSPECT}}}
\newcommand{\uv}{{\small {UV}}}
\newcommand{\dmB}{$\Delta m_{15}(B)$}

\newcommand{\agn}{{\small {AGN}}}

\newcommand{\het}{{\small {HET}}}

\newcommand{\mlcs}{{\small {MLCS}}}
\newcommand{\salt}{{\small {SALT}}}
\newcommand{\saltii}{{\small {SALT-II}}}

\newcommand{\signoi}{{\small {S/N}}}

\newcommand{\ugriz}{{\it {ugriz}}}

\newcommand{\pdf}{{\small {PDF}}}

\newcommand{\lrg}{{\small {LRG}}}
\newcommand{\rms}{{\small {RMS}}}
\newcommand{\cc}{{\small {CC}}}

\newcommand\pia{$P_{\mathrm{Ia}}$}

\newcommand\pzia{$P_{z,{\mathrm{Ia}}}$}

\newcommand\chisq{$\chi^2$}
\newcommand\chir{$\chi^2_{r}$}
\newcommand\chizr{$\chi^2_{z,r}$}
\newcommand\av{$A_V$}
\newcommand\tmax{$T_{\mathrm{max}}$}
\newcommand\zspec{$z_{\mathrm{spec}}$}
\newcommand\zsdss{$z_{\mathrm{SDSS}}$}
\newcommand\zlc{$z_{\mathrm{lc}}$}
\newcommand\zphoto{$z_{\mathrm{photo}}$}
\newcommand\zhost{$z_{\mathrm{host}}$}

\newcommand\wia{\mathcal{W}_{\rm{Ia}}^{\rm{false}}}
\newcommand\niatrue{\mathcal{N}_{\rm{Ia}}^{\rm{true}}}
\newcommand\niafalse{\mathcal{N}_{\rm{Ia}}^{\rm{false}}}
\newcommand\niatot{\mathcal{N}_{\rm{Ia}}^{\rm{TOT}}}
\newcommand\niacut{\mathcal{N}_{\rm{Ia}}^{\rm{CUT}}}
\newcommand\effia{\epsilon_{\rm{Ia}}}
\newcommand\puria{\eta_{\rm{Ia}}}
\newcommand\fomia{\mathcal{C}_{\rm{FoM-Ia}}}
\newcommand\contia{\kappa_{\rm{Ia}}}

\newcommand\wzia{\mathcal{W}_{\rm{z,Ia}}^{\rm{false}}}
\newcommand\nziatrue{\mathcal{N}_{\rm{z,Ia}}^{\rm{true}}}
\newcommand\nziafalse{\mathcal{N}_{\rm{z,Ia}}^{\rm{false}}}

\newcommand\nziacut{\mathcal{N}_{\rm{z,Ia}}^{\rm{CUT}}}
\newcommand\effzia{\epsilon_{\rm{z,Ia}}}
\newcommand\purzia{\eta_{\rm{z,Ia}}}
\newcommand\fomzia{\mathcal{C}_{\rm{z,FoM-Ia}}}

\newcommand\psnid{{\small{PSNID}}}
\newcommand\snana{{\small{SNANA}}}
\newcommand\snphotcc{{\sc{\small{SNP}}hot{\small CC}}}
\newcommand\smp{{\sc{smp}}}
\newcommand\soft{{\small{SOFT}}}
\newcommand{\mcmc}{{\small {MCMC}}}

\newcommand\des{{\small DES}}
\newcommand\lsst{{\small LSST}}
\newcommand\panstarrs{Pan-{\small STARRS}}

\newcommand\lcdm{{\small $\Lambda$CDM}}

\begin{document}

\title{Photometric SN~Ia Candidates from the Three-Year SDSS-II SN Survey
  Data}

\author{
Masao~Sako\altaffilmark{1},
Bruce~Bassett\altaffilmark{2,3},
Brian~Connolly\altaffilmark{1},
Benjamin~Dilday\altaffilmark{4,5,6},
Heather~Cambell\altaffilmark{7},
Joshua~A.~Frieman\altaffilmark{8,9,10},
Larry~Gladney\altaffilmark{1},
Richard~Kessler\altaffilmark{8,9},
Hubert Lampeitl\altaffilmark{7},
John~Marriner\altaffilmark{10},
Ramon~Miquel\altaffilmark{11,12},
Robert~C.~Nichol\altaffilmark{7},
Donald~P.~Schneider\altaffilmark{13},
Mathew~Smith\altaffilmark{14},
and Jesper~Sollerman\altaffilmark{15}
}

\altaffiltext{1}{
  Department of Physics and Astronomy,
  University of Pennsylvania, 209 South 33rd Street,
  Philadelphia, PA 19104; masao@sas.upenn.edu
}

\altaffiltext{2}{
  South African Astronomical Observatory,
  P.O. Box 9,
  Observatory 7935, South Africa
}

\altaffiltext{3}{
  Department of Mathematics and Applied Mathematics,
  University of Cape Town,
  Rondebosch, South Africa 7701
}

\altaffiltext{4}{
  Las Cumbres Observatory Global Telescope Network,
  6740 Cortona Dr., Suite 102
  Goleta, CA 93117, USA
}

\altaffiltext{5}{
  Department of Physics,
  University of California, Santa Barbara,
  Broida Hall, Mail Code 9530,
  Santa Barbara, CA 93106-9530, USA
}

\altaffiltext{6}{
  Department of Physics \& Astronomy,
  Rutgers, the State University of New Jersey,
  136 Frelinghuysen Road,
  Piscataway, NJ 08854, USA 
}

\altaffiltext{7}{
  Institute of Cosmology and Gravitation,
  Dennis Sciama Building,
  University of Portsmouth,
  Portsmouth, PO1 3FX, UK 
}

\altaffiltext{8}{
  Department of Astronomy and Astrophysics,
  The University of Chicago,
  5640 South Ellis Avenue, Chicago, IL 60637, USA
}

\altaffiltext{9}{
  Kavli Institute for Cosmological Physics,
  The University of Chicago,
  5640 South Ellis Avenue
  Chicago, IL 60637, USA
}

\altaffiltext{10}{
  Center for Particle Astrophysics,
  Fermi National Accelerator Laboratory,
  P.O. Box 500, Batavia, IL 60510, USA
}

\altaffiltext{11}{
  Institut de F\'{i}sica d'Altes Energies,
  Universitat Aut\'{o}noma de Barcelona,
  E-08193 Barcelona, Spain
}

\altaffiltext{12}{
  Instituci\'{o} Catalana de Recerca i Estudis Avan\c{c}ats,
  E-08010 Barcelona, Spain 
}

\altaffiltext{13}{
 Department of Astronomy \& Astrophysics,
 The Pennsylvania State University,
 University Park, PA 16802, USA
}

\altaffiltext{14}{
 Astrophysics, Cosmology and Gravity Centre (ACGC),
 Department of Mathematics and Applied Mathematics,
 University of Cape Town,
 Rondebosch, South Africa 7700
}

\altaffiltext{15}{
 The Oskar Klein Centre
 Department of Astronomy, Albaova
 SE-106 91 Stockholm, Sweden
}



\received{April 13, 2011}
\accepted{June 14, 2011}

\shorttitle{Photometric SN Classification}
\shortauthors{Sako et al.}

\begin{abstract}

  We analyze the three-year \sdssii\ Superernova (\sn) Survey data and
  identify a sample of $1070$ photometric \snia\ candidates based on their
  multi-band light curve data.  This sample consists of \sn\ candidates with
  no spectroscopic confirmation, with a subset of 210 candidates having
  spectroscopic redshifts of their host galaxies measured, while the remaining
  860 candidates are purely photometric in their identification.  We describe
  a method for estimating the efficiency and purity of photometric
  \snia\ classification when spectroscopic confirmation of only a limited
  sample is available, and demonstrate that \snia\ candidates from
  \sdssii\ can be identified photometrically with $\sim 91\%$ efficiency and
  with a contamination of $\sim 6\%$.  Although this is the largest uniform
  sample of \sn\ candidates to date for studying photometric identification,
  we find that a larger spectroscopic sample of contaminating sources is
  required to obtain a better characterization of the background events.  A
  Hubble diagram using \sn\ candidates with no spectroscopic confirmation, but
  with host galaxy spectroscopic redshifts, yields a distance modulus
  dispersion that is only $\sim 20 - 40$\% larger than that of the
  spectroscopically-confirmed \snia\ sample alone with no significant bias.  A
  Hubble diagram with purely photometric classification and redshift-distance
  measurements, however, exhibit biases that require further investigation for
  precision cosmology.

\end{abstract}



\section{Introduction}
\label{section_intro}

  Measurements of luminosity distances to nearby Type~Ia Supervova (\snia)
  \citep{phillips93, hamuy96a} and their distant counterparts have played a
  central role in modern cosmology and the remarkable discovery of an
  accelerating universe \citep{riess98, perlmutter99}.  Many dedicated
  supernova (\sn) surveys and follow-up programs have since then acquired
  light curves and spectra for several thousands of \sn\ in various redshift
  ranges: 1) at $z \la 0.1$ by the Lick Observatory Supernova Search
  \citep{filippenko01, ganeshalingam10}, the CfA monitoring campaign
  \citep{riess99, jha06a, matheson08, hicken09}, SNFactory \citep{bailey09},
  Carnegie Supernova Project Low-$z$ Program \citep{contreras09, folatelli10},
  the Palomar Transient Factory \citep{rau09, law09}, and the Panoramic Survey
  Telescope and Rapid Response System
  (\panstarrs\footnote{\texttt{http://pan-starrs.ifa.hawaii.edu/public}}); 2)
  the \sdssii\ \sn\ Survey in the intermediate redshift interval $0.1 \la z
  \la 0.3$ \citep{frieman08, sako08}; 3) the highest-redshift range observable
  from the ground at $0.3 \la z \la 1$ by the Supernova Legacy Survey
  \citep[\snls;][]{astier06, guy10, conley11}, the \essence\ \sn\ Survey
  \citep{miknaitis07, wood-vasey07}, the Carnegie Supernova Project High-$z$
  Program \citep{freedman09}; and finally 4) $z \ga 1$ \snia\ from space using
  the Hubble Space Telescope \citep{riess04a, riess07, dawson09}.

  Many future surveys, such as the Dark Energy Survey
  \citep[\des;][]{flaugher10} and the Large Synoptic Survey Telescope
  \citep[\lsst;][]{lsst09}, with deeper and more wide-field imaging
  capabilities will probe much larger volumes of the universe allowing
  discoveries of thousands to tens of thousands of high-redshift
  \sn\ candidates each year.  Spectroscopic follow-up observations of these
  large, faint \sn\ samples will require prohibitively large time allocations
  with existing instruments.  Studies of \sn\ properties and cosmology will,
  therefore, necessitate a photometric determination of the \sn\ type,
  cosmological redshift, and the luminosity distance from light curves with
  possibly a limited subsample with spectroscopic confirmation and redshift
  measurements.

  Various methods for photometrically classifying \sn\ have been discussed in
  the literature.  Optical and \uv\ colors near maximum light, for example,
  have been used to distinguish \snia\ from core-collapse
  \sn\ \citep{pskovskii77, poznanski02, panagia03, riess04b, johnson06}.
  \citet{poznanski07a} have developed a Bayesian method that classifies
  \sn\ using only a single epoch of photometry \citep[see
    also,][]{kuznetsova07, rodney09}.  Template-fitting methods have been
  employed for spectroscopic targetting of active \sn\ candidates
  \citep{sullivan06, sako08}.  \citet{sullivan06} have performed an analysis
  to identify a sample of photometric \snia\ candidates from the first year of
  the Supernova Legacy Survey.  \citet{dahlen04}, \citet{poznanski07b},
  \citet{dahlen08}, \citet{dilday08}, \citet{dilday10}, \citet{rodney10b}, and
  \citet{graur11} have also used photometric classification to measure
  \sn\ rates as a function of redshift.

  Although an efficient photometric \sn\ classifier is crucial for a
  successful spectroscopic follow-up program and also for understanding the
  bias in the spectroscopic sample, the ability to estimate both the
  efficiency and purity of the selected sample is also important for
  understanding, for example, possible biases in distance measurements and
  studies of \sn\ rates.  Clearly, the efficiency can be improved by
  compromising purity, and vice versa, and the requirements may vary depending
  on the type of study involved.

  In addition to photometrically identifying \snia\ candidates, redshifts as
  well as luminosity distances can be inferred from the same multi-band light
  curve data.  These studies of {\it \sn\ cosmology without spectroscopy} have
  been pioneered by \cite{barris04} and carried out more recently by a number
  of authors.  \citet{palanque10}, \citet{kessler10a}, and \citet{rodney10a}
  for example, study the quality of photometric redshifts on large samples of
  existing data.  \citet{rodney10a} also construct a photometry-only Hubble
  diagram of the first-year \sdssii\ and \snls\ spectroscopically-confirmed
  \snia\ using their Supernova Ontology with Fuzzy Templates (\soft) method.
  Others show comparisons of measured and input redshifts primarily from
  simulations \citep{kim07, kunz07, wang07a, wang07b, gong09, scolnic09}.

  The accuracy and precision of the measured parameters depend on many
  observational factors including the statistical quality of the observed
  light curves, surface brightness of the underlying host galaxy, photometric
  calibration, wavelength coverage, the number of filter bandpasses, and the
  observing cadence.  Other non-observational factors that might affect the
  measurements are the quality of the light curve models, assumptions on the
  dust properties and intrinsic \sn\ colors, as well as priors used in the
  fits.  The photometric redshift uncertainty on any individual \sn\ is
  obviously larger than a typical spectroscopic redshift error, but a
  substantially larger number of {\emph{unbiased}} redshift and distance
  measurements made possible photometrically might be able to provide
  competitive constraints on cosmological parameters with future large-scale
  surveys.

  Some of the existing softwares and algorithms, including the one presented
  in this paper, were recently used to participate in the Supernova
  Photometric Classification Challenge \citep{kessler10b}, a public
  competition for classifying \sn\ light curves.  The authors of the challenge
  released a large number of simulated \sn\ light curves of undisclosed types
  and a small ``spectroscopic'' sample with known redshifts and types for
  training.  Participants of the challenge submitted their classifications as
  well as photometric redshifts if available.  The algorithm presented here
  achieved the highest overall figure of merit, though there is significant
  room for improvement.

  This paper focuses on understanding these issues using an improved
  implementation of existing methods and through analysis of a much larger
  sample of \sn\ candidates for testing.  We use the three-year
  \sdssii\ \sn\ Survey data as our test bed to identify photometric
  \snia\ candidates with realistic estimates of sample purity.  The
  description of the photometric classification algorithm and the
  spectroscopic and photometric \sn\ samples from \sdssii\ are presented in
  \S\ref{sec:sncand}.  The procedures for estimating the \snia\ typing
  efficiency and purity using the spectroscopic sample are described
  \S\ref{sec:fom} and \S\ref{sec:eff}.  The properties of the photometric
  \snia\ candidates identified are described in \S\ref{sec:photo-ia}.  The
  quality of the light curve photometric redshifts is discussed in
  \S\ref{sec:redshift}.  Comparisons with simulations are shown in
  \S\ref{sec:sims}.  Finally, our results are summarized in
  \S\ref{sec:summary}.

\section{The SDSS-II SN Candidates}
\label{sec:sncand}

  The \sdssii\ \sn\ Survey was conducted during the September -- November
  months of 2005 -- 2007.  A 300~deg$^2$ region along the celestial equator
  was observed using the \sdss\ 2.5m telescope \citep{gunn98, fukugita96,
    york00, gunn06} with an average cadence of four days \citep{frieman08,
    dr7}.  The survey depth and area are optimal for discovering and measuring
  light curves of \snia\ at intermediate redshifts ($0.1 \la z \la 0.4$),
  complementing other surveys.  During the search campaigns, new variable and
  transient sources detected in the difference images were designated as
  ``\sn\ candidates''.  After each night of imaging observations on the
  \sdss\ telescope, the \sn\ candidates were photometrically classified based
  on the available multiband light curves, and a subset of the events were
  observed spectroscopically close to their moment of discovery
  \citep{sako08}.  Photometry and results from follow-up spectroscopy from the
  first season are presented in \citet{holtzman08} and \citet{zheng08},
  respectively, and measurements of the cosmological parameters from the
  first-year sample and studies of the sources of systematic uncertainties are
  presented in \citet{kessler09a}, \citet{sollerman09}, and
  \citet{lampeitl10}.

  Over 10000 \sn\ candidates were discovered during the three-year
  \sdssii\ \sn\ Survey, and the majority of these candidates are
  spectroscopically unconfirmed due to limited spectroscopic resources.  The
  goal of this paper is to photometrically identify the \snia\ candidates, and
  to estimate the efficiency and purity of that photometric classification.
  We investigate whether reliable cosmological measurements can be performed
  from \sn\ candidates without spectroscopic confirmation.  We first describe
  the \sn\ classification algorithm below, and then discuss our method for
  estimating the efficiency and purity using a limited number of
  spectroscopically-confirmed \sn.

\subsection{Photometric SN Classification Algorithm}
\label{subsec:psnid}

  The candidates are classified using a light curve analysis software called
  ``Photometric \sn\ IDentification'' (\psnid), which is an extended version
  of the software used for prioritizing spectroscopic follow-up observations
  for the \sdssii\ \sn\ Survey as described in \citet{sako08}\footnote{The
    software is included in the \snana\ Package \citep{snana}.  A standalone
    version is also available directly from the author.}.  Extensive tests
  were performed using the publicly-available \snana\ light curve
  simulations\footnote{\texttt{http://sdssdp62.fnal.gov/sdsssn/SIMGEN\_PUBLIC/}}
  as well as the data presented here.  \psnid\ was also used to analyze
  simulations from the Supernova Photometric Classification Challenge and
  achieved the highest overall figure of merit \citet[hereafter
    K10b]{kessler10b}.  Briefly, the software uses the observed photometry,
  calculates the reduced \chisq\ (\chir~$= \chi^2$ per degree of freedom)
  against a grid of \snia\ light curve models and core-collapse \sn\ (\ccsn)
  templates, and identifies the best-matching \sn\ type and set of parameters
  with, and without, host galaxy redshift as priors in the grid search.  A
  number of important improvements have been made, which are described below.

\begin{deluxetable}{lllr}
  \tablewidth{0pt}
  \tablecaption{Core-collapse \sn\ Templates\label{tbl:cc_templates}}
  \tablehead{
    \colhead{Type} &
    \colhead{Subtype} &
    \colhead{IAU Name} &
    \colhead{SDSS ID}
  }
  \startdata
  Ibc     & Ib   & SN2005hl & 2000 \\
  \nodata & Ib   & SN2005hm & 2744 \\
  \nodata & Ic   & SN2006fo & 13195 \\
  \nodata & Ib   & SN2006jo & 14492 \\
  II      & II-L/P & SN2004hx & 18 \\
  \nodata & II-P & SN2005lc & 1472 \\
  \nodata & II-P & SN2005gi & 3818 \\
  \nodata & II-P & SN2006jl & 14599 \\
  \enddata
\end{deluxetable}

\begin{figure*}[htb]
  \begin{center}
    \includegraphics[angle=-90, width=.4\textwidth]{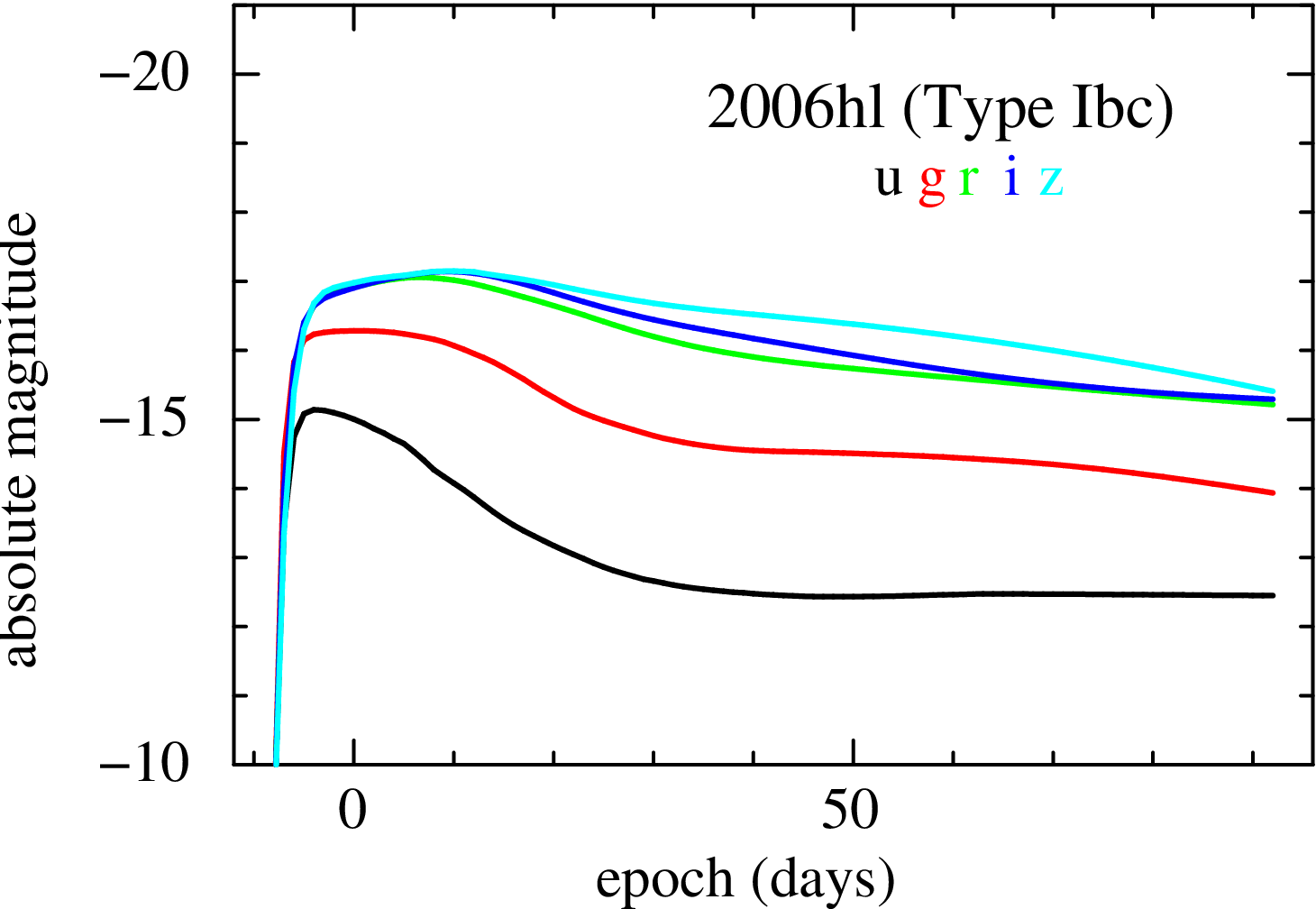}
    \includegraphics[angle=-90, width=.4\textwidth]{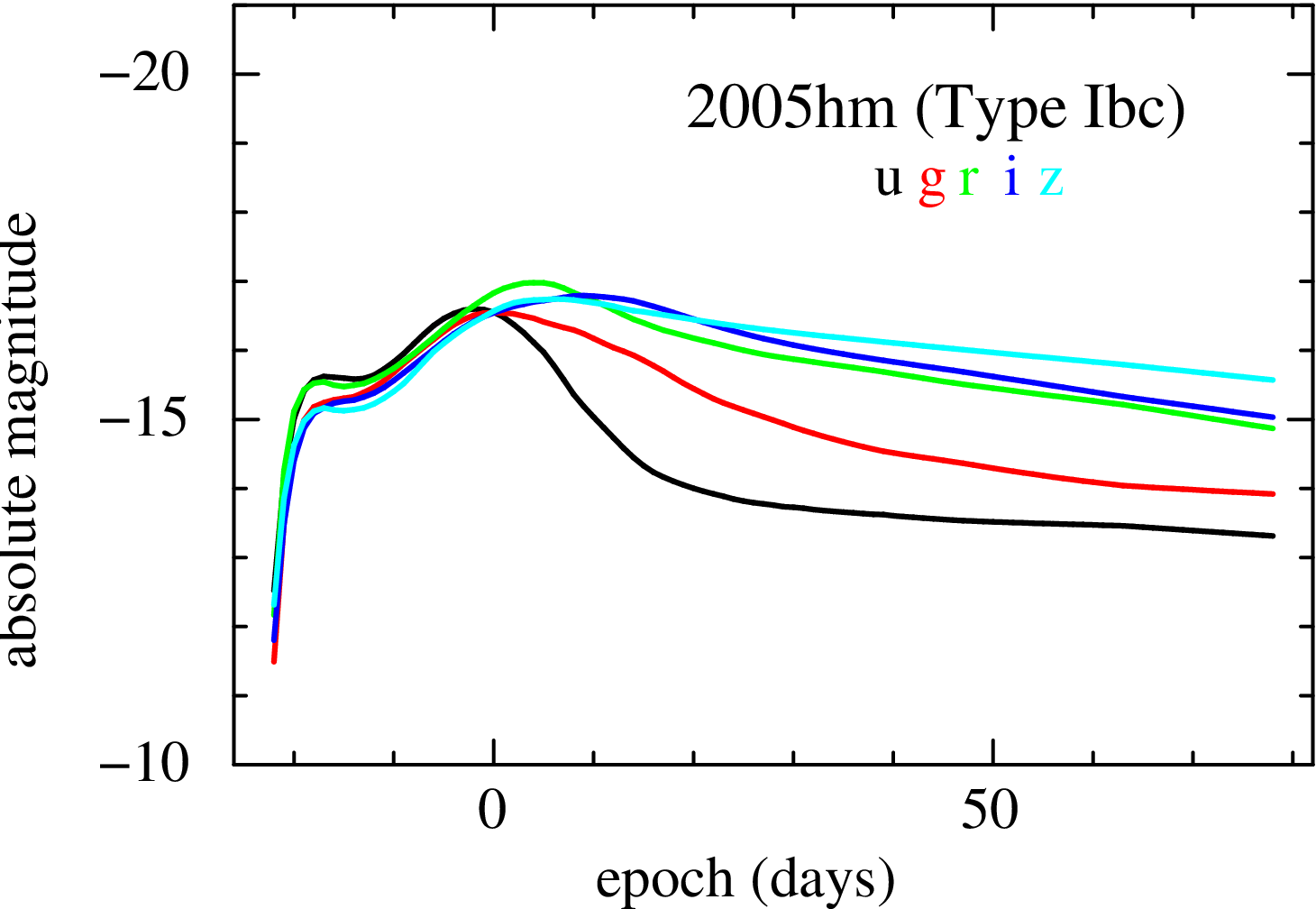}
    \includegraphics[angle=-90, width=.4\textwidth]{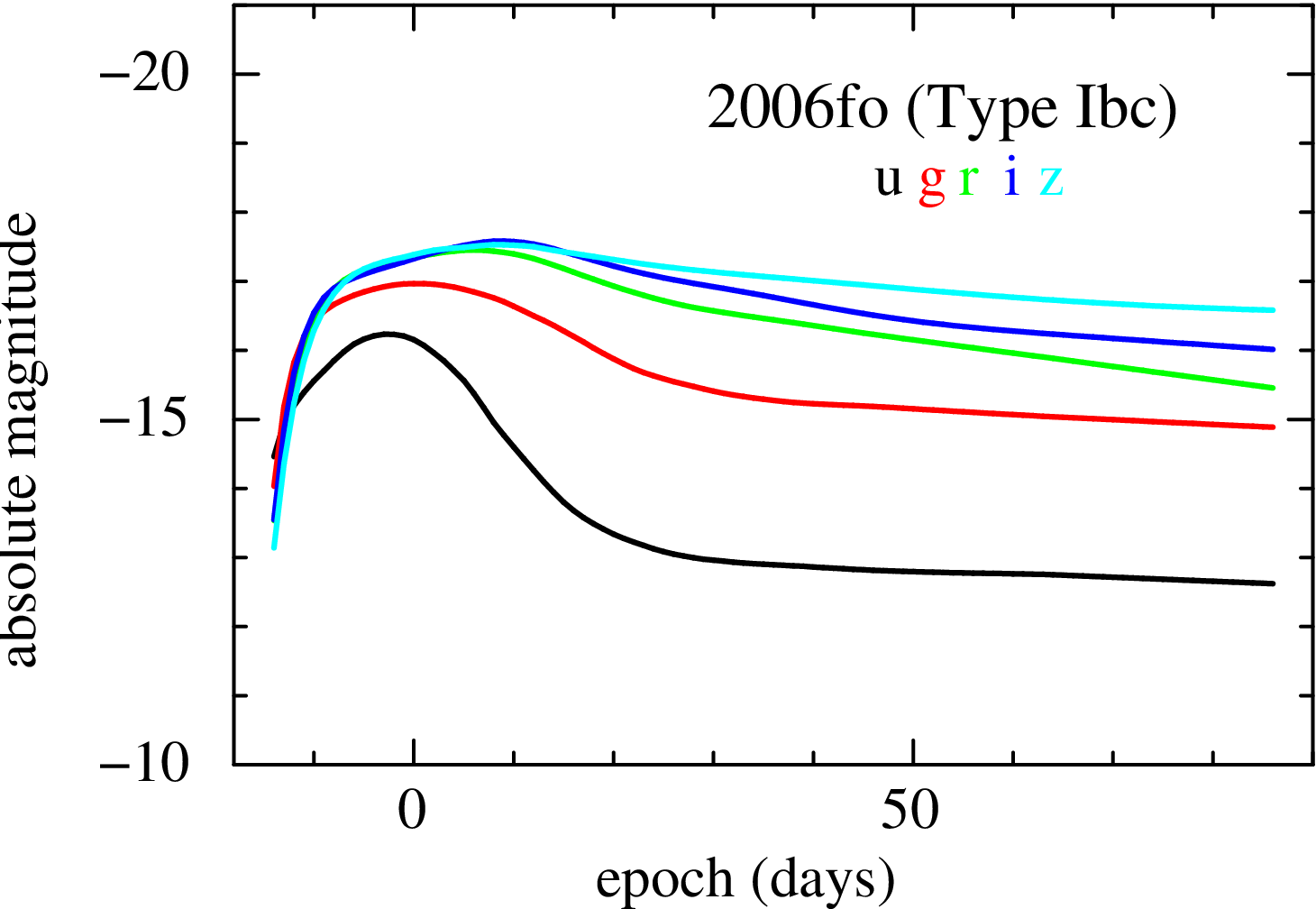}
    \includegraphics[angle=-90, width=.4\textwidth]{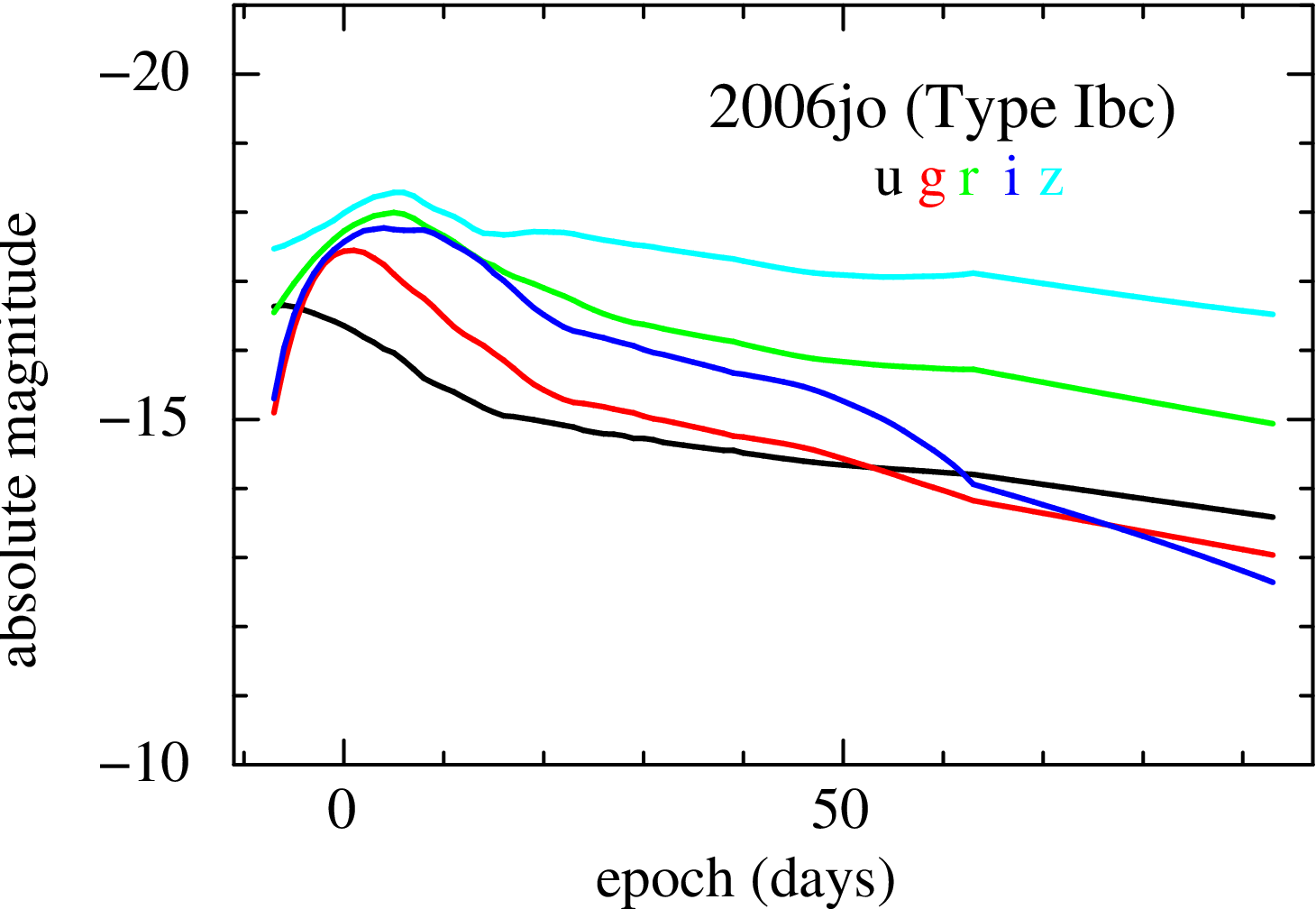}
  \end{center}
  \caption{Absolute magnitude light curves of \snibc\ discovered and observed
    by \sdssii, which are part of the template library -- \sn\ 2005hl (top
    left), \sn\ 2005hm (top right), \sn\ 2006fo (bottom left), and \sn\ 2006jo
    (bottom right).  [See online version for color figures.]}
  \label{fig:snibc_templates}
\end{figure*}

\begin{figure*}[htb]
  \begin{center}
    \includegraphics[angle=-90, width=.4\textwidth]{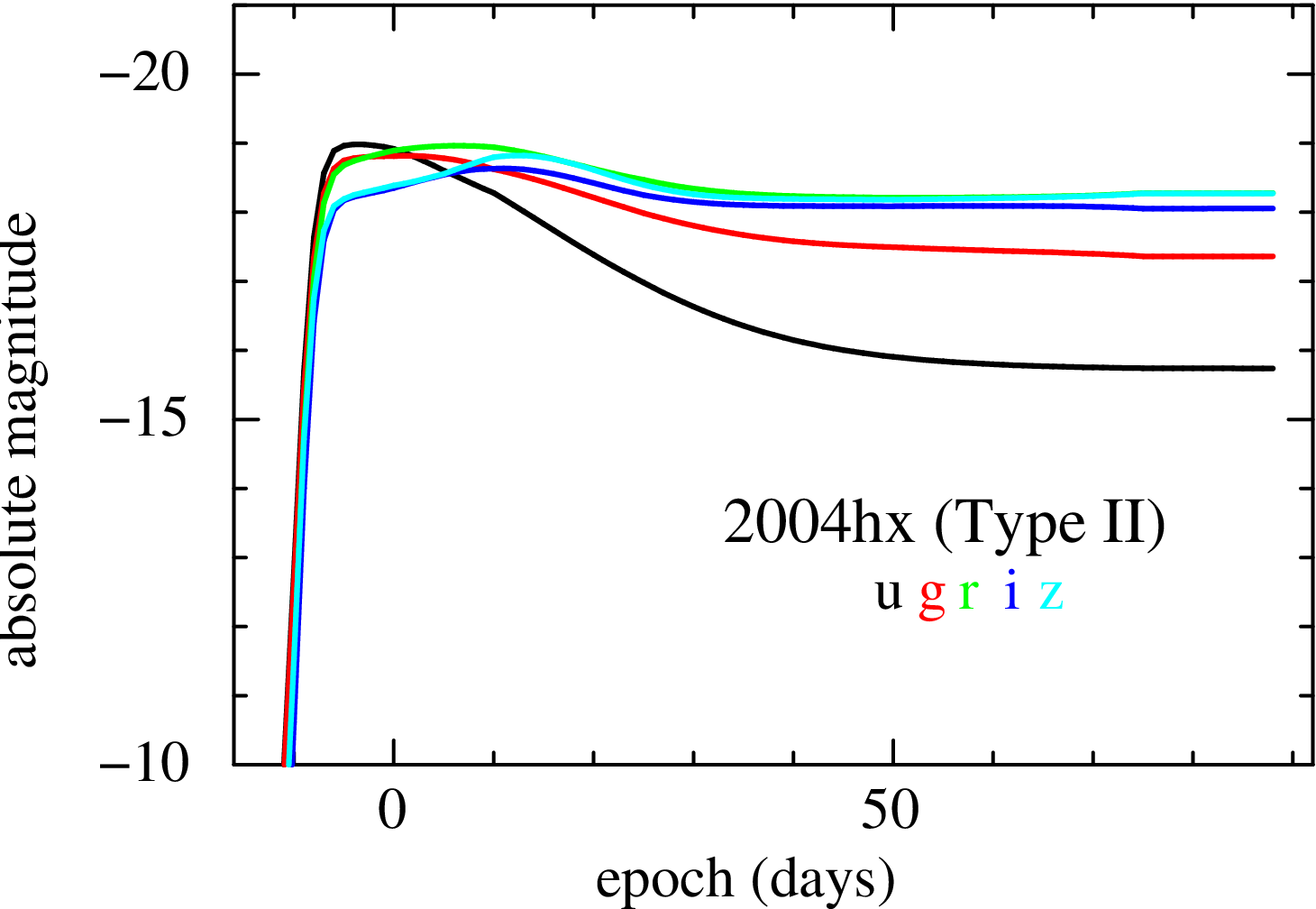}
    \includegraphics[angle=-90, width=.4\textwidth]{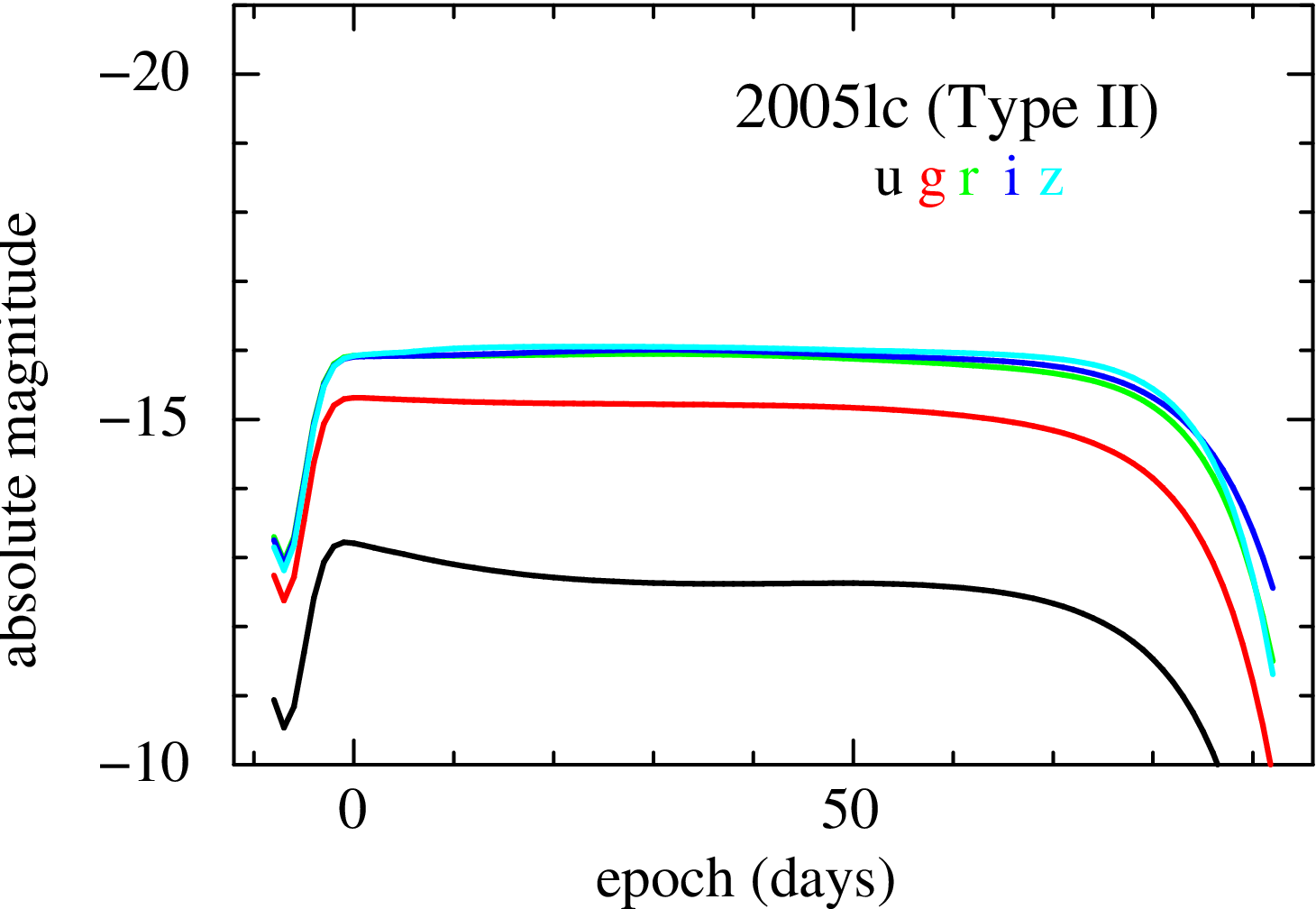}
    \includegraphics[angle=-90, width=.4\textwidth]{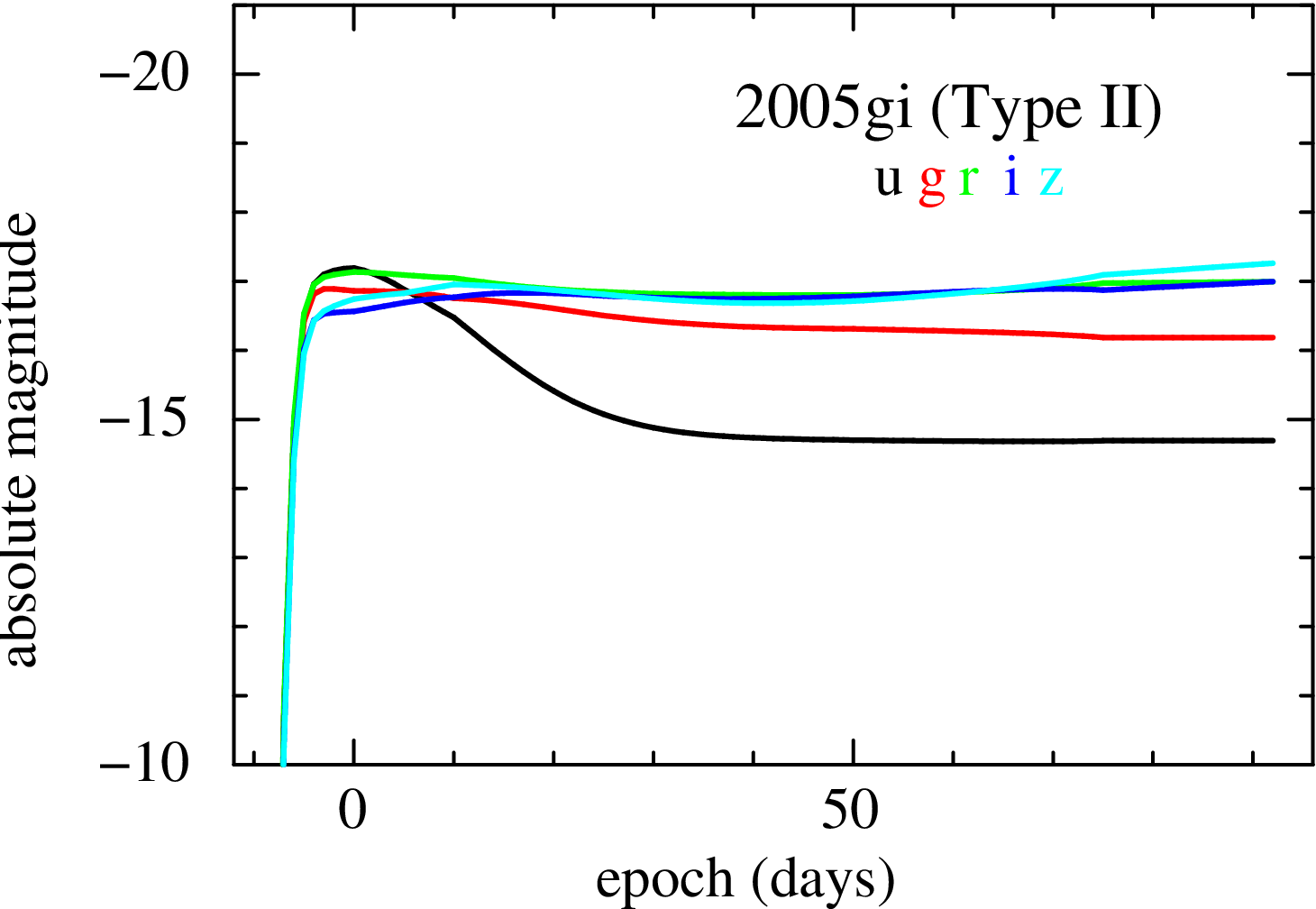}
    \includegraphics[angle=-90, width=.4\textwidth]{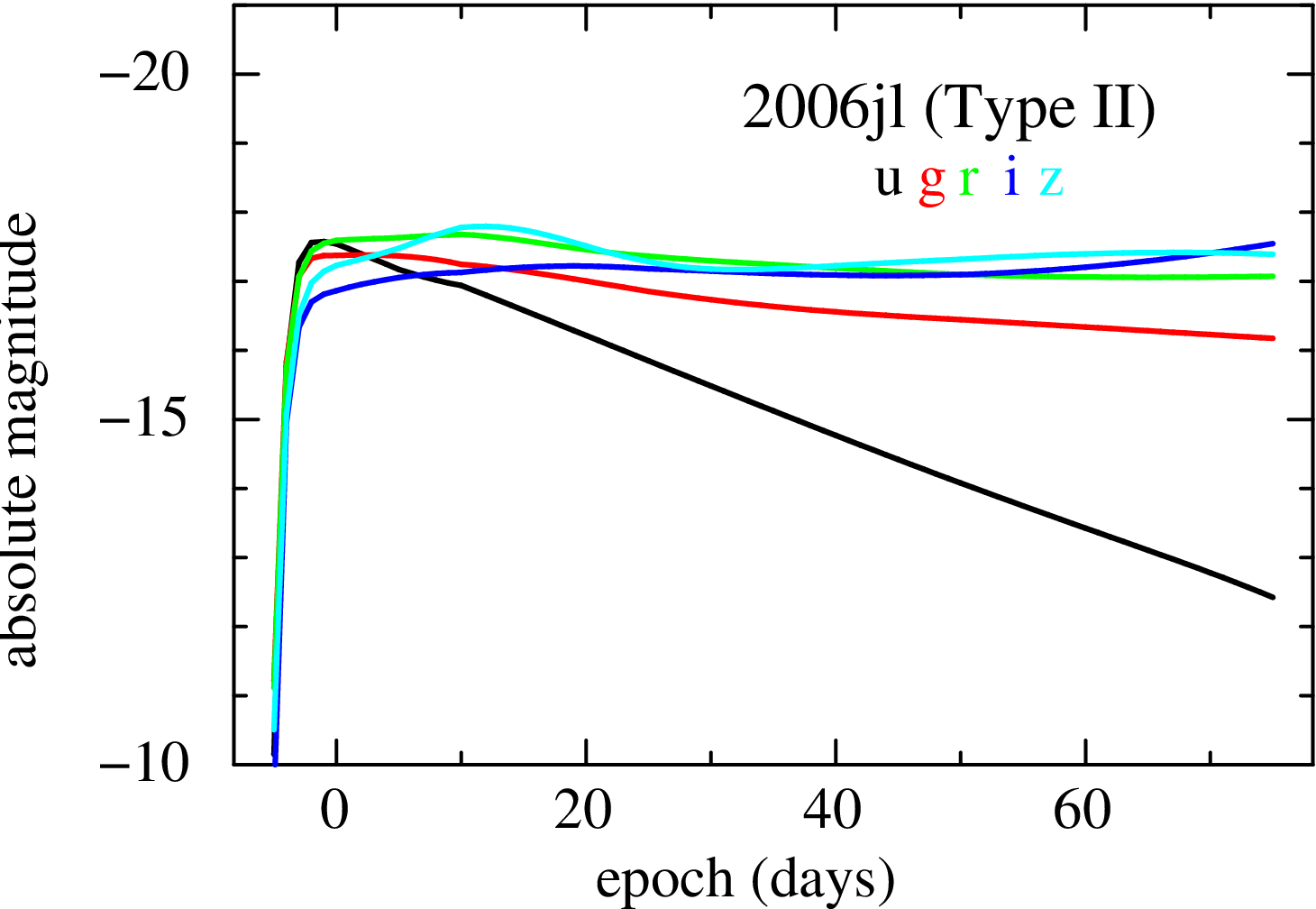}
  \end{center}
  \caption{Absolute magnitude light curves of \snii\ discovered and observed
    by \sdssii, which are part of the template library -- \sn\ 2004hx (top
    left), \sn\ 2005lc (top right), \sn\ 2005gi (bottom left), and \sn\ 2006jl
    (bottom right).  [See online version for color figures.]}
  \label{fig:snii_templates}
\end{figure*}

  First, in addition to finding the light curve model with the minimum
  \chir\ through a grid search, the software computes the Bayesian
  probabilities that a candidate could be a Type Ia, Type Ib/c, or a Type II
  \sn.  The algorithm is similar to that of \citet{poznanski07a} except that we
  subclassify \ccsn\ into Types Ib/c and II using an extended set of templates
  (see below), and also allow the \snia\ light curve shape parameter and
  distance modulus to vary in the fits.  Specifically, we calculate the
  Bayesian Evidence $E$ by marginalizing the product of the likelihood
  function and prior probabilities over the model parameter space.  For the
  \snia\ models, there are five model parameters -- redshift $z$, $V$-band
  host galaxy extinction \av, time of maximum light \tmax,
  \dmB\ \citep{phillips93, phillips99}, and distance modulus $\mu$.  Milky Way
  extinction is modeled assuming the \citet{cardelli87} law with $R_V = 3.1$,
  while extinction in the \sn\ host galaxy assumes a total-to-selective
  extinction ratio of $R_V \equiv A_V/E(B-V)= 2.2$ \citep{kessler09a}.  Priors
  in \av, \tmax, and $\mu$ can also be applied optionally, but we set them to
  be flat in this present work.  For the redshift, we evaluate each light
  curve twice using 1) a flat prior and 2) a gaussian prior if an external
  redshift estimate $z_{\mathrm{ext}}$ and uncertainty $\sigma_z$ are
  available from either the host galaxy (photometric or spectroscopic
  redshift) or the \sn\ spectrum.  The \snia\ Bayesian evidence is therefore,
\begin{equation}
  E_{\mathrm{Ia}} = \int_{\mathrm{all~ parameters}}
  P(z)~e^{-\chi^2/2}~dz~dA_V~dT_{\mathrm{max}}~d\Delta m_{15,B}~d\mu,
\end{equation}
  where,
\begin{equation}
  P(z) = \frac{1}{\sqrt{2\pi\sigma_z}}~e^{-(z-z_{\mathrm{ext}})^2/2\sigma_z^2}.
\end{equation}
  When an external redshift is not available, we assume the prior to be flat
  by setting $P(z) = 1$.  For the \snibc\ and \snii\ models, the integral over
  \dmB\ is replaced with a summation over the individual templates used in the
  comparison,
\begin{equation}
  E_{\mathrm{Ibc,II}} = \sum_{\mathrm{templates}} \int
  P(z)~e^{-\chi^2/2}~dz~dA_V~dT_{\mathrm{max}}~d\mu.
\end{equation}
  The Bayesian probability of one of the three possible \sn\ types is then
  given by,
\begin{equation}
  P_{\mathrm{type}} =
  \frac{E_{\mathrm{type}}}{E_{\mathrm{Ia}}+E_{\mathrm{Ibc}}+E_{\mathrm{II}}}.
\end{equation}
  The probabilities $P_{\mathrm{type}}$ and minimum \chir\ values calculated
  using the gaussian spectroscopic redshift prior are denoted with a subscript
  $z$ (i.e., $P_{\mathrm{z,type}}$ and \chizr).  External photometric
  redshifts of the host galaxies are not used in the fits in this work.  The
  probabilities are normalized such that,
\begin{equation}
  P_{\mathrm{Ia}} + P_{\mathrm{Ibc}} + P_{\mathrm{II}} = 1,
\end{equation}
  which is equivalent to assuming that the \sn\ candidate is a real \sn\ and
  not another class of variable sources.  This assumption is reasonable, since
  sources in Stripe~82 with a prior history of variability and other
  multi-year variables are rejected from our analysis \citep{sako08}.  This
  set of Bayesian probabilities is useful because it quantifies the relative
  likelihood of \sn\ types -- the best-fit minimum \chir\ alone is not a good
  indicator of the most likey \sn\ type.  As advocated by
  \citet{kuznetsova07}, we therefore select \snia\ based on both the Bayesian
  probability \pia\ and the goodness-of-fit \chir.

  Next, although the \snia\ light curve models used herein are the same as
  those described in \citet{sako08}, we have assigned empirical model errors
  that yield reasonable \chir\ values for light curves with high
  \signoi\ ratio.  The assumed magnitude errors $\delta m$ on the $gri$ model
  light curves depend on the rest-frame epoch $t$ in days from $B$-band
  maximum as follows,
\begin{equation}
  \delta m_{\rm{Ia}} = \left\{ \begin{array}{ll}
    0.08 + 0.04 \times (|t|/20) & |t| < 20~\mathrm{days}, \\
    0.12 + 0.08 \times ((|t|-20)/60) & |t| \ge 20~\mathrm{days}.
  \end{array} \right.
\end{equation}
  The \ccsn\ light curve templates have error in $gri$ given by,
\begin{equation}
  \delta m_{\rm{CC}} = 0.08 + 0.08 \times (|t|/60)
\end{equation}
  for all epoch.  The model errors in $u$ and $z$ are chosen to be twice the
  above values due to larger intrinsic model variations and calibration
  uncertainties in these bands.  These $\delta m$ parameters were determined
  to provide reasonable \chir\ values (\chir~$\sim 1$) primarily for nearby
  \sn\ candidates with small photometric errors.  They do not affect the fit
  results of faint candidates.

  Third, we adopt \ccsn\ light curve templates from a sample of nearby
  \sn\ discovered and observed by \sdssii.  Specifically, we use four
  \snibc\ templates and four \snii\ templates as listed in
  Table~\ref{tbl:cc_templates}.  The \sdssii\ \ccsn\ light curve templates
  were generated using the \citet{nugent02} spectral templates, interpolating
  between epochs, and warping them to match each of the observed \ugriz\ light
  curves at their respective spectroscopic redshifts.  For all \snibc, we use
  Nugent's normal Ibc spectral templates, and we use the Type II-P templates
  for all \snii.  The \snii\ light curve photometry are available from
  \citet{dandrea10}.

  The set of eight core-collapse templates listed in
  Table~\ref{tbl:cc_templates} were selected from a larger group of 24
  templates (5 Nugent, 11 SDSS-II, and 8 from the
  \suspect\footnote{\texttt{http://bruford.nhn.ou.edu/$\sim$suspect/index1.html}}
  database) by empirically maximizing the purity of the confirmed
  \snia\ sample.  Core-collapse templates that either frequently misidentify
  \snia\ as \ccsn\ or correctly identify only a small number of confirmed
  \ccsn\ were excluded.  Rare, peculiar \snia\ are also excluded from our
  template library.  We also do not include templates for other types of
  variable sources, most notably the active galactic nuclei (\agn), since
  there are other ways of rejecting the majority of these events.  The
  rest-frame absolute magnitude \ugriz\ light curves of the eight \ccsn\ used
  as templates in this analysis are shown in Figures~\ref{fig:snibc_templates}
  and \ref{fig:snii_templates}.

  Finally, while the Bayesian classification probabilities are computed
  through marginalization over the grid of the model parameters, the posterior
  probability distributions for each of the five parameters are estimated by
  running a Markov Chain Monte Carlo (\mcmc).  This results in a significant
  reduction of computing time and more reliable estimates of the parameter
  uncertainties, since the probability distributions are often asymmetric,
  show significant correlations, and can often have more than one local
  maximum.  It is also straightforward to incorporate additional model
  parameters and priors.

\begin{figure*}[tb]
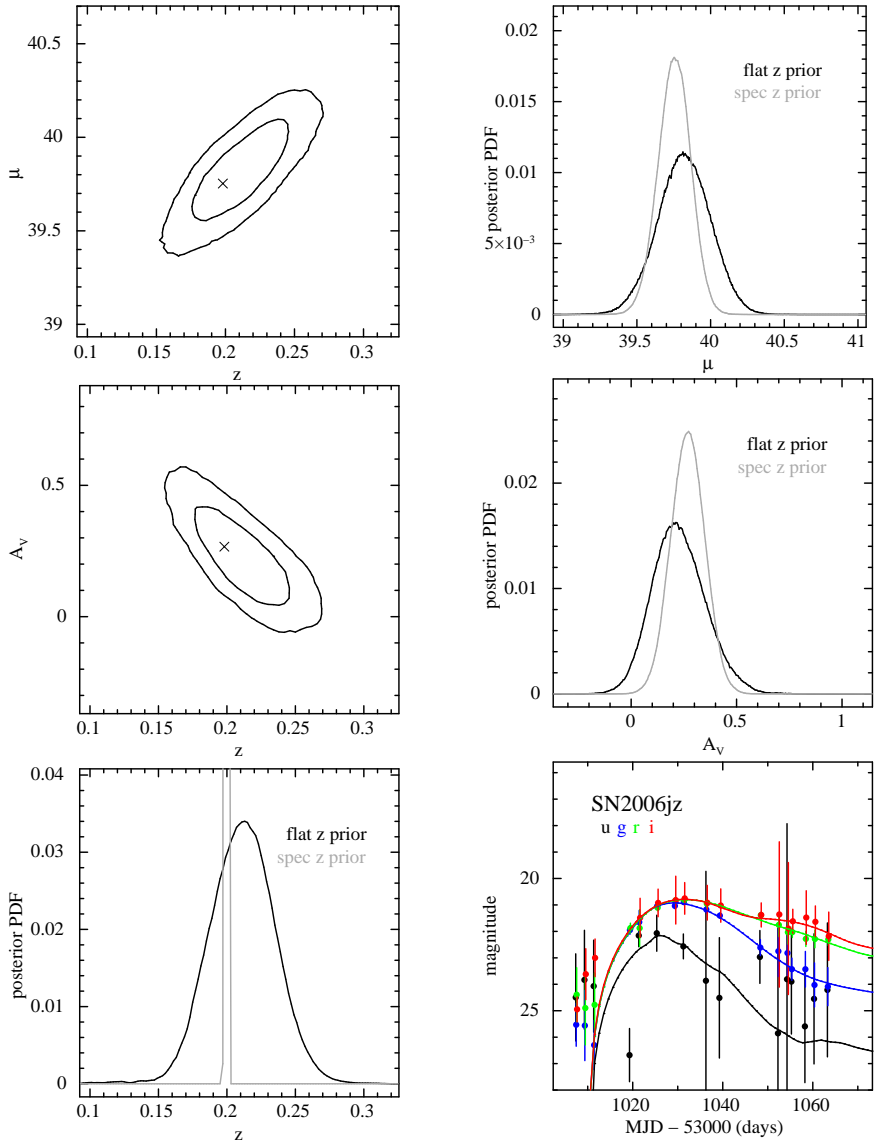

  \begin{center}
  \begin{minipage}[t]{2.4in}
    \includegraphics[angle=-90, width=0.85\textwidth]{fig3a.ps}
    \includegraphics[angle=-90, width=0.85\textwidth]{fig3b.ps}
    \includegraphics[angle=-90, width=0.85\textwidth]{fig3c.ps}
  \end{minipage}
  \begin{minipage}[t]{2.4in}
    \includegraphics[angle=-90, width=0.85\textwidth]{fig3d.ps}
    \includegraphics[angle=-90, width=0.85\textwidth]{fig3e.ps}
    \includegraphics[angle=-90, width=0.85\textwidth]{fig3f.ps}
  \end{minipage}
  \end{center}
  \caption{An example of the posterior probability distribution functions
    (\pdf) for a spectroscopically-confirmed \snia\ 2006jz at $z = 0.20$.  The
    observed $ugri$ light curves and the best-fit \snia\ model are shown on
    the bottom right panel.  The top-left and middle-left panels show 1- and
    2-$\sigma$ contours in the $z$-$\mu$ and $z$-\av\ planes, respectively,
    assuming a flat redshift prior.  The $\times$ indicates the median
    parameter values when a spectroscopic-redshift prior is used.  The two
    panels on the right and the bottom-left panel show the posterior \pdf\ in
    $\mu$, \av, and $z$ marginalized over the other 4 parameters using the
    flat (black) and spectroscopic (gray) redshift priors.}
  \label{fig:mcmc}
\end{figure*}

\begin{figure*}[tb]
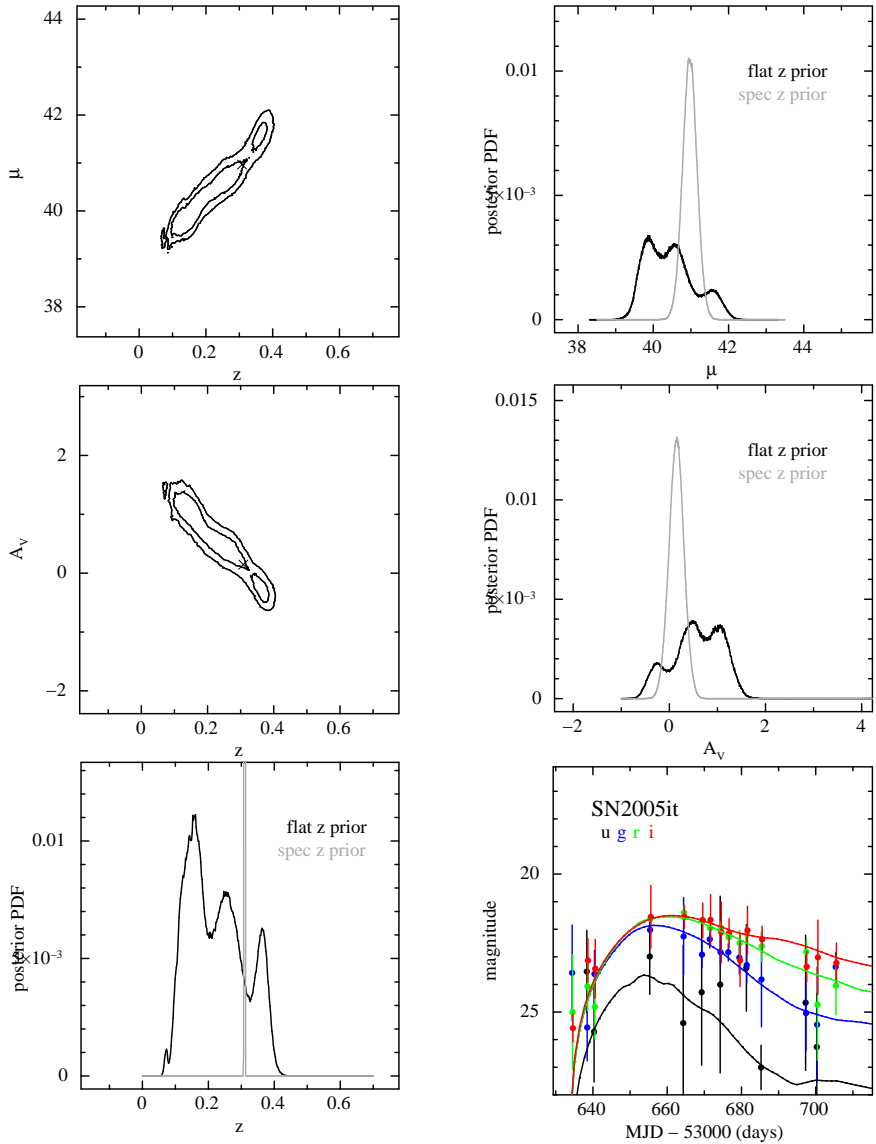

  \begin{center}
  \begin{minipage}[t]{2.4in}
    \includegraphics[angle=-90, width=0.85\textwidth]{fig4a.ps}
    \includegraphics[angle=-90, width=0.85\textwidth]{fig4b.ps}
    \includegraphics[angle=-90, width=0.85\textwidth]{fig4c.ps}
  \end{minipage}
  \begin{minipage}[t]{2.4in}
    \includegraphics[angle=-90, width=0.85\textwidth]{fig4d.ps}
    \includegraphics[angle=-90, width=0.85\textwidth]{fig4e.ps}
    \includegraphics[angle=-90, width=0.85\textwidth]{fig4f.ps}
  \end{minipage}
  \end{center}
  \caption{Same as in Figure~\ref{fig:mcmc} for a spectroscopically-confirmed
    \snia\ 2005it at $z = 0.30$.}
  \label{fig:mcmc2}
\end{figure*}

  Figure~\ref{fig:mcmc} shows an example output from \psnid\ for a
  spectroscopically-confirmed \snia, 2006jz at $z=0.20$.  Derived parameter
  constraints from the \mcmc\ are shown for both the flat and spectroscopic
  redshift priors.  There are two general points that are worth noting.
  First, $z$ and \av\ are anti-correlated in the sense that a low-$z$,
  high-\av\ \snia\ is similar to a high-$z$, low-\av\ event.  This is
  expected, since redshift and dust both have the effect of reddening the
  light curves.  But since dust also attenuates the light, a larger \av\ value
  must be compensated for by putting the event at a smaller distance modulus.
  This happens in the way such that $z$ and $\mu$, marginalized over the other
  three parameters, are positively correlated.  The slope of this correlation
  is redshift-dependent.  Second, the widths of the marginalized $\mu$ and
  \av\ probability distribution function (\pdf) for the flat redshift prior
  are only a factor $\sim 2$ larger than those for a spectroscopic redshift
  prior.  This general behavior is true for most of our well-observed \snia,
  although the constraints using a flat-$z$ prior degrades dramatically at
  higher redshifts, as shown in Figure~\ref{fig:mcmc2} for a $z=0.30$
  confirmed \snia\ 2005it.

\subsection{Confirmed and Unconfirmed Samples}
\label{sec:samples}

  We first divide the full sample of \sn\ candidates into two groups -- the
  spectroscopically confirmed and unconfirmed samples.  The unconfirmed sample
  consists of sources of unknown type with no spectroscopy of the active
  \sn\ candidate, but a subset of the events do have spectroscopy of their
  host galaxies.  The spectroscopically-confirmed sample consists of \snia,
  \snibc, \snii, as well as variable \agn.  This sample is used to study the
  classification criteria and also allows us to estimate the selection
  efficiency and purity, which is a crucial part of our analysis.  The $ugriz$
  multi-band light curves of all \sn\ candidates are constructed using the
  Scene-Modeling Photometry method \citep[\smp;][]{holtzman08} and analyzed
  using the \psnid\ software described above.

  The full \sn\ sample is analyzed with \psnid, and we select the candidates
  that have light curve coverage and signal-to-noise (\signoi) ratio that are
  appropriate for photometric \snia\ classification.  Specifically, we
  consider only the candidates that meet the following three criteria: (1)
  Have at least one epoch of photometry near peak at $-5 < t < +5$~days in the
  \sn\ rest frame and at least one additional epoch after peak at $t >
  +15$~days, which are determined from to the best-fit \snia\ model,
  irrespective of whether or not the fit is acceptable; (2) Have maximum
  \signoi\ ratio greater than five in at least two of the $gri$ bands, and;
  (3) Were detected during only one search season.  These cuts are referred to
  as the light curve quality cuts.

  The spectroscopically-confirmed sample consists of 508 \snia, 80 \ccsn\ (18
  \snibc, 62 \snii), and 202 \agn\footnote{Of the 202 \agn, 58 are in the
    {\small{DR}}7 spectroscopic quasar catalog from \citet{schneider10}.}.  We
  refer to these as the ``conf-Ia'', ``conf-\cc'', and the ``conf-\agn''
  samples.  After imposing the light curve quality cuts, this sample is
  reduced to 367 \snia, 45 \ccsn, and 83 \agn, for a total of 495 events when
  a flat spectroscopic redshift prior is used.  Using the spectroscopic
  redshift prior results in 551 events.  The numbers differ since the two
  forms of the redshift priors can result in best-fit \snia\ models with
  dramatically different dates of maximum light, especially for the \agn.

  There is a significant bias in the spectroscopically-confirmed \sn\ sample
  toward brighter events.  For the \sdssii\ \sn\ Survey, our primary goal was
  to discover and study the properties of \snia, so only a small fraction of
  \ccsn\ candidates were observed for spectroscopy.  A detailed study of the
  impact on photometric \snia\ typing due to contaminating sources is,
  therefore, limited by this small number of spectroscopically confirmed
  \ccsn.

\begin{figure}[htb]
  \begin{center}
    \includegraphics[angle=-90, width=.45\textwidth]{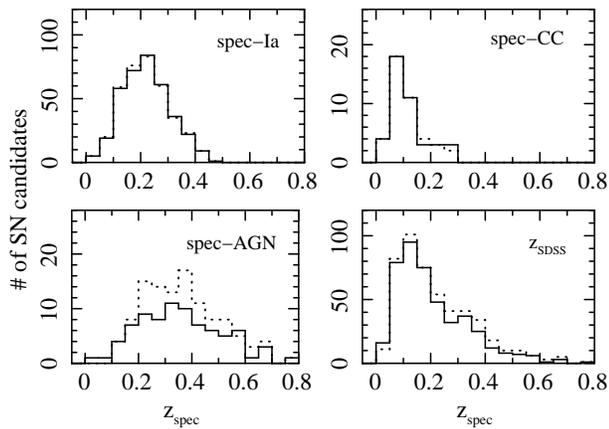}
  \end{center}
  \caption{The redshift distributions of the conf-Ia (top left),
    conf-\cc\ (top right), conf-\agn\ (bottom left), and \zsdss\ (bottom
    right) samples used in our studies (see \S~\ref{sec:samples} for
    descriptions).  The solid and dashed histograms represent the samples that
    pass our light curve quality cuts with the flat and spectroscopic redshift
    priors, respectively.  The redshift bins are $\Delta z = 0.05$ wide.}
  \label{fig:spec_zdist}
\end{figure}

  To help quantify this bias, we identified the \sn\ candidates that are
  associated with galaxies with spectra from the \sdss\ spectroscopic survey
  \citep{eisenstein01, strauss02, richards02}.  These galaxies have
  well-defined selection criteria and, as we describe below, will help
  quantify the spectroscopic targeting bias and to obtain a better estimate of
  the level of contamination from non-\snia\ events.  There are a total of
  2369 \sn\ candidates that are within $10\arcsec$ from an
  \sdss\ spectroscopic galaxy.  This sample is referred to as the ``\zsdss''
  sample.  After light curve quality cuts, there are 448 and 499 sources for
  the flat and spectroscopic redshift priors, respectively, which includes
  both confirmed and unconfirmed \sn\ candidates.  The majority of the sources
  are rejected because of their multi-year variability, suggesting that these
  sources are likely variable \agn\ whose nuclear activity is not immediately
  apparent from their optical spectra.  The samples are summarized in
  Table~\ref{tbl:spectro}.  The redshift distributions of the four different
  spectroscopic samples are shown in Figure~\ref{fig:spec_zdist}.

\begin{deluxetable*}{lrrrrrrr}
  \tabletypesize{\footnotesize}
  \tablewidth{0pt}
  \tablecaption{The \sdssii\ Spectroscopic Sample\label{tbl:spectro}}
  \tablehead{
     & & \multicolumn{3}{c}{Flat Redshift Prior} &
    \multicolumn{3}{c}{Spectroscopic Redshift Prior} \\
    \colhead{Type} &
    \colhead{Total} &
    \colhead{Good\tablenotemark{a}} &
    \colhead{\pia\ $\geq 0.9$} &
    \colhead{\pia\ $\leq 0.1$} &
    \colhead{Good\tablenotemark{a}} &
    \colhead{\pia\ $\geq 0.9$} &
    \colhead{\pia\ $\leq 0.1$}
  }
  \startdata
  Confirmed \snia\  & 508 & 367 & 357 &   2 & 371 & 366 &   1 \\
  Confirmed \ccsn\  &  80 &  45 &  14 &  30 &  45 &  11 &  32 \\
  Confirmed \agn\   & 202 &  83 &  32 &  44 & 135 &  86 &  46 \\
  \hline
  \sn\ with $z_{\rm{SDSS}}$  & 2369 & 448 & 248 & 159 & 499 & 317 & 150 \\
  \hline  \hline
  Total            & 3159 & 732 & 539 & 201 & 788 & 599 & 163 \\
  \enddata

  \tablenotetext{a}{This sample includes \sn\ that satisfy the following
    photometric quality criteria: (1) There is at least one epoch of
    photometry at $-5 < t < +5$~days from peak and another epoch at $+5 < t <
    +15$~days from peak for the best-fit \snia\ model; (2) There is at least
    two filter measurements with \signoi~$>5$; (3) The candidate was detected
    in only a single search season.}

\end{deluxetable*}

  The unconfirmed sample consists of a total of 3221 candidates that pass the
  same light curve quality cuts.  Of these 3221 candidates, 2776 have no
  spectroscopic observations, while the remaining 445 candidates are either
  part of the \zsdss\ sample described above (230 candidates) or have host
  galaxy redshifts from our own follow-up observations (215 candidates).

\begin{figure}[htb]
  \begin{center}
    \includegraphics[angle=-90, width=.45\textwidth]{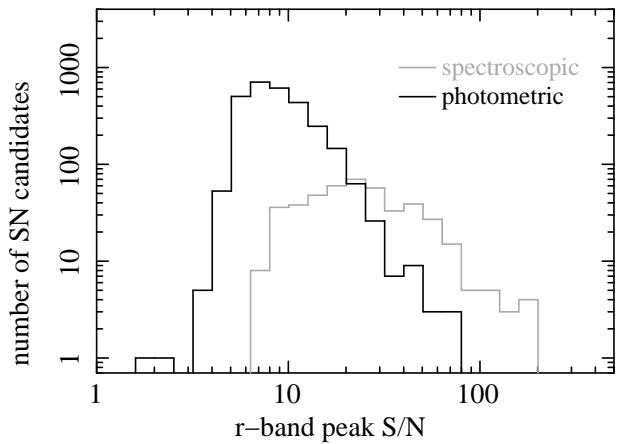}
  \end{center}
  \caption{The distributions of maximum $r$-band signal-to-noise ratio
    (\signoi) of the spectroscopically-confirmed \sn\ candidates (dashed) and
    photometric candidates (solid) considered in this work.  The spectroscopic
    sample has an average peak \signoi\ of $\sim 30$ while the photometric
    sample has average \signoi\ of $\sim 10$.}
  \label{fig:snr}
\end{figure}

  A histogram of the maximum $r$-band \signoi\ of this sample is shown in
  Figure~\ref{fig:snr}.  The mean \signoi\ of $\sim 30$ for the spectroscopic
  sample is substantially higher than that of the photometric sample, which
  has a mean \signoi\ of $\sim 10$.  The implications of this difference are
  discussed in \S~\ref{sec:summary}.

\section{SN Classification Figure of Merit}
\label{sec:fom}

  Since our goal here is to identify \snia, we define the photometric typing
  efficiency $\effia$ as the fraction of \snia, after software \signoi\ light
  curve quality cuts, that are photometrically identified as \snia.  Letting
  $\niatrue$ be the number of true \snia\ photometrically identified as
  \snia\ and $\niacut$ be the total number of \snia\ in the sample after the
  light curve quality cuts, we define the photometric \snia\ selection
  efficiency to be,
\begin{equation}
  \label{equ:eff}
  \effia = \frac{\niatrue}{\niacut}.
\end{equation}
  Note that this is not the true \snia\ identification efficiency since the
  denominator $\niacut$ includes only the events that pass the \signoi\ and
  light curve quality cuts.  In terms of the total number of
  \snia\ ($\niatot$) that were detected in the area observed by the survey,
\begin{equation}
  \label{equ:cutvstot}
  \niacut = \epsilon_{\rm{CUT}} \niatot
\end{equation}
  where $\epsilon_{\rm{CUT}}$ is, in general, a function of $z$, $A_V$, \dmB,
  peak magnitude, time of maximum light, software detection threshold,
  requirements on light curve \signoi\ and temporal coverage, as well as the
  observing conditions.  The determination of the value of
  $\epsilon_{\rm{CUT}}$ is beyond the scope of the paper, but the effect of
  our selection cuts can be modeled using the \snana\ Package.

  Adopting the convention similar to that used in evaluating the
  \sn\ Photometric Classification Challenge (hereafter \snphotcc; K10b) we
  define the {\it{photometric purity}} $\puria$ as the fraction of the
  candidates identified as \snia\ that are actual \snia\ with a penalty factor
  $\wia$ described below.  Letting $\niafalse$ be the number of
  non-\snia\ incorrectly identified as \snia, the photometric purity of the
  sample is,
\begin{equation}
  \label{equ:pur}
  \puria = \frac{\niatrue}{\niatrue + \sum_i
    \mathcal{W}_{\rm{Ia},i}^{\rm{false}} \mathcal{N}_{\rm{Ia},i}^{\rm{false}}},
\end{equation}
  where the sum in the denominator allows for several classes $i$ of
  contaminating sources (e.g., \ccsn, \agn, and variable stars) possibly with
  different penalty factors.  We define a figure of merit ($\fomia$) as,
\begin{equation}
  \label{equ:fom}
  \fomia = \effia \times \puria.
\end{equation}
  This definition of $\fomia$ is designed for real data and differs from the
  {\it pseudo-purity} from the \snphotcc\ by the unknown factor
  $1/\epsilon_{\rm{CUT}}$, i.e., $\fomia =
  \fomia^{\rm{SNPhotCC}}/\epsilon_{\rm{CUT}}$.  K10b also define the {\it true
    purity} to be the case for $\wia = 1$.  This figure of merit is only one
  measure of success, and it is not necessarily the optimal measure for all
  types of studies.  Higher \snia\ purity might be more important than
  efficiency for certain studies, and vice versa.  Finally, we define the
  {\it{contamination}} $\contia$ as,
\begin{equation}
  \label{equ:contami}
  \contia = 1 - \puria.
\end{equation}
  These quantities determined with the spectroscopic redshift prior are
  designated with a subscript $z$.

  To give a simple numerical example, consider a survey that is capable of
  detecting 100 \snia\ that pass \signoi\ and light curve quality cuts.  A
  photometric classifier that identifies 90 candidates as \snia, of which 10
  are actually non-Ia events has an efficiency of $\effia = 80/100 = 0.80$,
  purity of $\puria = 80/90 = 0.89$, and contamination of $\contia = 1 - 0.89
  = 0.11$.  In practice, however, these quantities can be determined only for
  the spectroscopically confirmed \sn\ sample for which the correct type is
  known.  The efficiency, purity, or some combination of these two parameters
  can be optimized by choosing the appropriate values for \pia\ and \chir.  If
  the spectroscopic sample is an unbiased representation of all of the
  \sn\ candidates, then one can expect the efficiency and the purity of both
  the spectroscopic and photometric samples to be the same within statistical
  uncertainties.  However, this is almost never the case in a blind
  \sn\ survey given limited spectroscopic resources.  \sn\ candidates that are
  brighter and/or suffer less host galaxy contamination will have higher
  spectroscopic success and completeness.  This is illustrated in
  Figure~\ref{fig:snr}, which shows that the light curve peak \signoi\ of the
  spectroscopic sample is on average a factor of $\sim 3$ higher than that of
  the photometric sample.  Below we describe a method to correct for this bias
  and to estimate the efficiency and purity of the photometric sample using a
  limited and biased spectroscopic training set.

\section{Estimating the Efficiency and Purity}
\label{sec:eff}

\subsection{SN~Ia Identification With Spectroscopic Redshifts}
\label{subsec:withz}

\begin{figure}[tb]
  \begin{center}
    \includegraphics[angle=0, width=.40\textwidth]{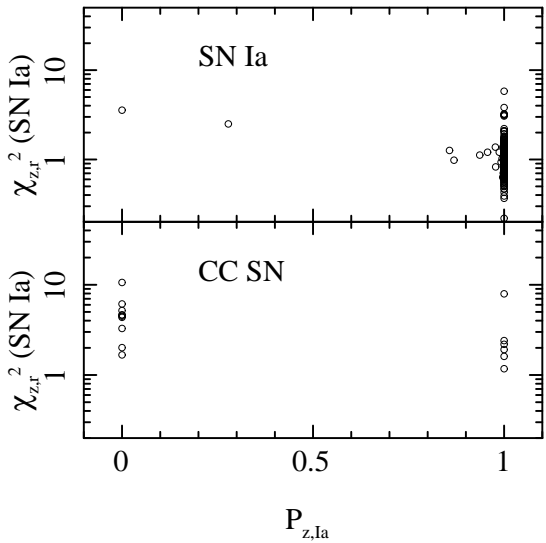}
  \end{center}
  \caption{The distribution of \pzia\ and \chizr\ values for the
    spectroscopically-confirmed \snia\ (top panel) and \ccsn.  Spectroscopic
    redshifts are used as priors in all of the fits.}
  \label{fig:pzia_chisq}
\end{figure}

\begin{figure}[tb]
  \begin{center}
    \includegraphics[angle=0, width=.45\textwidth]{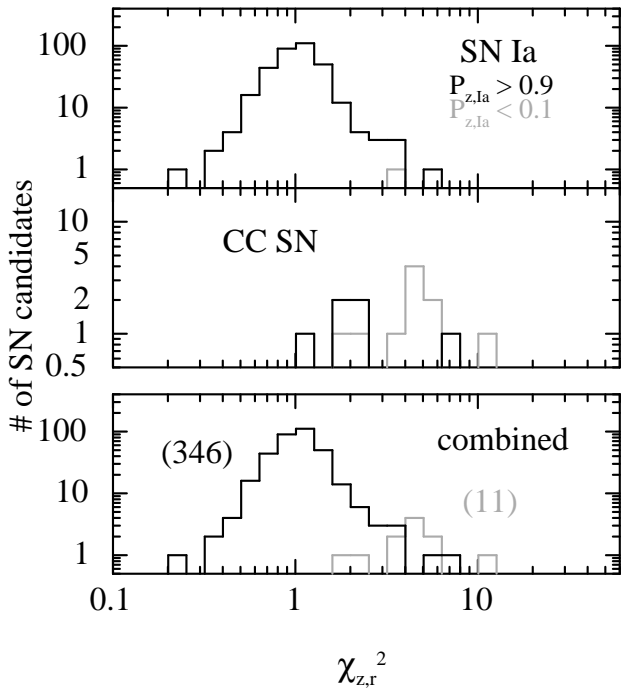}
  \end{center}
  \caption{Histograms of best-fit \chizr\ values for a \snia\ model for
    $P_{\mathrm{Ia}} \ge 0.9$ (black) and $P_{\mathrm{Ia}} \le 0.1$ (gray) for
    the spectroscopically confirmed \snia\ (top panel) and \ccsn\ (bottom
    panel).}
  \label{fig:zchisq_hist}
\end{figure}

  We first estimate the efficiency and purity of photometric
  \snia\ identification when spectroscopic redshifts are used as priors in the
  light curve fits.  We determine $\nziatrue$ and $\nziafalse$ from the
  spectroscopic \snia\ and \ccsn\ and how they depend on the minimum
  \pzia\ and the maximum allowed \chizr.  This is relevant for future
  \sn\ surveys that will, for example, obtain spectra of all \sn\ candidate
  host galaxies after the search, but not spectra of all the active
  \sn\ candidates.  The values for \pzia\ and \chizr\ are shown in
  Figure~\ref{fig:pzia_chisq} separately for the spectroscopically confirmed
  \snia\ and \ccsn\ samples.

  As shown in the top panel of Figure~\ref{fig:pzia_chisq}, all but a handful
  of \snia\ are well fit to a \snia\ model.  Of the $\nziacut = 371$
  spectroscopic \snia\ that pass the light curve quality cuts, 366 sources
  have \pzia~$\geq 0.9$.  Only a single \snia\ (\sn\ 2007qd;
  \citealt{mcclelland10}) has \pzia~$\le 0.1$.  This event is a nearby
  peculiar 2002cx-like event, which is underluminous compared to normal
  \snia\ and has an extremely low expansion velocity \citep{li03, jha06b}.
  There are other nearby peculiar \snia\ in our sample (\sn\ 2005hk
  \citealt{phillips07}, \sn\ 2005gj \citealt{aldering06, prieto07}), but these
  candidates were detected over two search seasons due to their brightness and
  slow decline, and were, therefore, rejected.  The bottom panel of the same
  figure, however, shows that a substantial fraction of the spectroscopic
  \ccsn\ also satisfy \pzia~$\ge 0.9$ implying that the contamination can be
  significant depending on the maximum allowed \chizr\ value used for the
  \snia\ identification.  Specifically, 11 out of the 45 \ccsn\ (24\%) that
  satisfy our light curve quality cuts have \pzia~$\ge 0.9$.  If no other cuts
  are invoked, then $\nziatrue = 366$ and $\nziafalse = 11$.  We also note
  that the majority of the sources have either \pzia~$\sim 0$ or \pzia~$\sim
  1$, so both $\nziatrue$ and $\nziafalse$ are not sensitive to the precise
  choice of the minimum \pzia.

  Before determining how $\nziatrue$ and $\nziafalse$ depend on the choice of
  the maximum \chizr, we note that 5 of the 11 \ccsn\ with \pzia~$\ge 0.9$ can
  be rejected by requiring the light curve \photoz\ (\zlc), using a flat
  redshift prior, to be within $3\sigma$ of the spectroscopic redshift \zspec;
  i.e., $|$\zlc\ -- \zspec$|$/$\sigma_z < 3$.  We reject candidates that fail
  this cut, and show the distributions of the \chizr\ values for the
  \snia\ and \ccsn\ in Figure~\ref{fig:zchisq_hist} for \pzia~$\ge 0.9$ and
  \pzia~$\le 0.1$.  Of the 366 \snia\ and 11 \ccsn\ with good light curves and
  \pzia~$\geq 0.9$, 22 and 5 candidates, respectively, are rejected by this
  requirement on redshift agreement.  Therefore, there are only 6 \ccsn\ that
  satisfy all \snia\ selection cuts.

  In the last step, we estimate the unknown factor $\wzia$, which can be
  interpreted as a penalty factor for spectroscopic incompleteness and
  targeting biases.  The \sdssii\ \sn\ Survey follow-up strategy was to
  observe the ``good'' \snia\ candidates at higher priority than the
  \ccsn\ candidates, especially for the fainter ($r \ga 20.5$~mag) sources due
  to limited spectroscopic resources.  A simple interpretation of this factor
  is that if our follow-up strategy had instead been to observe a random
  sample of \sn\ candidates, then we would have spectroscopically identified
  $\wzia$ times more \ccsn.

  One way to estimate this bias factor is to select a subsample of
  \sn\ candidates with spectroscopic redshifts, which is representative of the
  underlying distribution of the \sn\ types.  The ratio of these candidates
  with \pzia~$\le 0.1$ to those with \pzia~$\ge 0.9$ can then be interpreted
  to be approximately the ratio of \ccsn\ to \snia\ in our survey.

\begin{figure}[htb]
  \begin{center}
    \includegraphics[angle=-90, width=.45\textwidth]{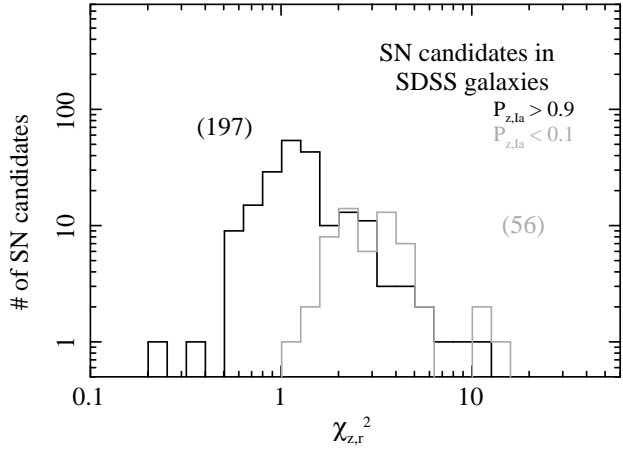}
  \end{center}
  \caption{Histograms of best-fit \chir\ values for a \snia\ model for
    $P_{\mathrm{z,Ia}} \ge 0.9$ (black) and $P_{\mathrm{z,Ia}} \le 0.1$ (gray)
    for the \sn\ candidates in \sdss\ spectroscopic galaxies using the
    redshift as a prior.}
  \label{fig:chizsq_hist_sdss}
\end{figure}

\begin{figure}[tb]
  \begin{center}
    \includegraphics[angle=-90, width=.45\textwidth]{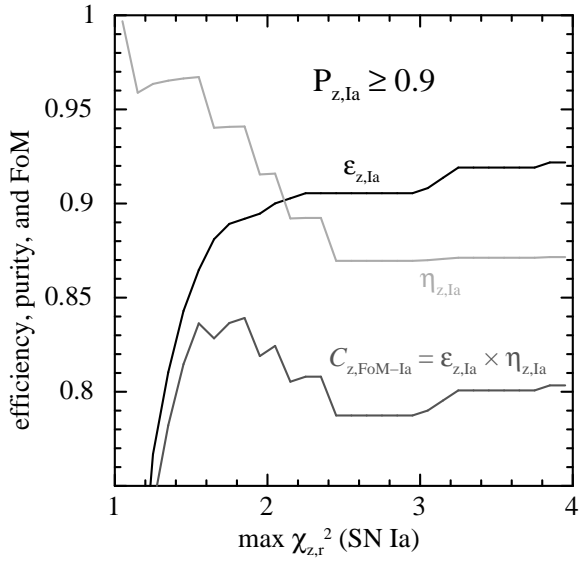}
  \end{center}
  \caption{The efficiency, purity, and figure of merit for the
    spectroscopically-confirmed \sn\ as functions of the maximum-allowed
    \chizr\ for \pzia~$\geq 0.9$.}
  \label{fig:zeff_pur_fom}
\end{figure}

\begin{figure}[thb]
  \begin{center}
    \includegraphics[angle=0, width=.40\textwidth]{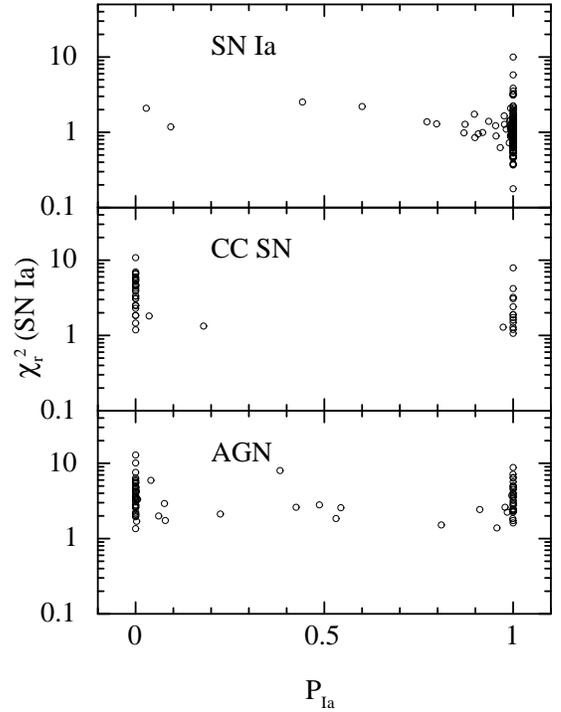}
  \end{center}
  \caption{The distributions of \pia\ and \chir\ values for the
    spectroscopically confirmed \snia\ (top panel), \ccsn\ (middle panel), and
    \agn\ (bottom panel).  The fits were performed with a flat redshift
    prior.}
  \label{fig:pia_chisq}
\end{figure}

  This can be done by considering the \sn\ candidates in galaxies with
  redshifts from the \sdss\ spectroscopic survey, which has a set of
  well-defined selection criteria.  We identify candidates in the main galaxy
  \citep{strauss02}, quasar \citep{richards02}, and the Luminous Red Galaxy
  (\lrg; \citealt{eisenstein01}) samples.  The \lrg\ sample is several
  magnitudes deeper than the main galaxy sample and consists primarily of
  passive galaxies with old stellar populations, which do not host any \ccsn.
  We include this sample to account for the fact that \snia\ are also on
  average a few magnitudes more luminous than \ccsn, so a magnitude-limited
  survey will discover many more \snia\ than \ccsn.  The distributions of
  \chizr\ for \pzia~$\geq 0.9$ and \pzia~$\leq 0.1$ for candidates in the
  \sdss\ galaxy spectroscopy sample with $|$\zlc\ -- \zspec$|$/$\sigma_z < 3$
  are shown in Figure~\ref{fig:chizsq_hist_sdss}.  The ratio of the number of
  candidates with \pzia~$\geq 0.9$ to those with \pzia~$\leq 0.1$ is $197/56 =
  3.5$ compared to $350/11 = 32$ for the combined spectroscopic sample shown
  in the bottom panel of Figure~\ref{fig:zchisq_hist}.  The bias (penalty)
  factor for the spectroscopic sample can, therefore, be estimated to be
  $\wzia = 32/3.5 = 9.0$.  An unbiased spectroscopic follow-up strategy would
  have resulted in $\wzia = 9.0$ times more contaminating \ccsn\ for
  \snia\ identification.

  We use this penalty factor to calculate $\effzia$ and $\purzia$ as functions
  of the maximum \chizr.  The expression for the purity is,
\begin{equation}
  \label{equ:zpur_spec}
  \purzia = \frac{\nziatrue}{\nziatrue + \mathcal{W}_{\rm{z,Ia}}^{\rm{false}}
    \mathcal{N}_{\rm{z,Ia}}^{\rm{false}}}.
\end{equation}
  Figure~\ref{fig:zeff_pur_fom} shows how $\effzia$, $\purzia$, and $\fomzia$
  depend on the maximum-allowed \chizr\ for \pzia~$\ge 0.9$.  The figure of
  merit has a broad maximum value of $\fomzia \sim 0.84$ at approximately
  \chizr~$= 1.8$, where the efficiency and purity are $\sim 89$\% and $\sim
  94$\%, respectively.  A caveat to the estimate of $\purzia$ is that it is
  based on only six confirmed \ccsn\ that pass our \snia\ selection cuts.

\subsection{SN~Ia Identification without Spectroscopic Redshifts}
\label{sec:withoutz}

  We next determine $\niatrue$ and $\niafalse$ when no external redshift
  information is available to provide additional constraints in the light
  curve fits.  Here we have an additional source of contaminating sources --
  variable \agn\ -- which can be identified if either the galaxy spectrum is
  available or the candidate is variable over a long period of time ($\ga
  1$~year).  We use the confirmed \sn\ and the \agn\ samples discussed in
  \S~\ref{sec:samples} to determine how the efficiency, purity, and figure of
  merit depend on the minimum \pia\ and the maximum allowed \chir\ using the
  flat redshift prior.  The three panels in Figure~\ref{fig:pia_chisq} show
  the \pia\ and \chir\ values for the spectroscopic \snia, \ccsn, and
  \agn\ samples.  As with the previous case, most of the spectroscopic
  \snia\ are clustered near \pia~$\sim 1$ and \chir~$\sim 1$ indicating that
  they are well-fit to \snia\ models.  There are also a handful of \ccsn\ and
  \agn\ with \pia~$\sim 1$, however, so the amount of contamination can again
  be substantial depending on the maximum allowed \chir.

\begin{figure}[thb]
  \begin{center}
    \includegraphics[angle=0, width=.40\textwidth]{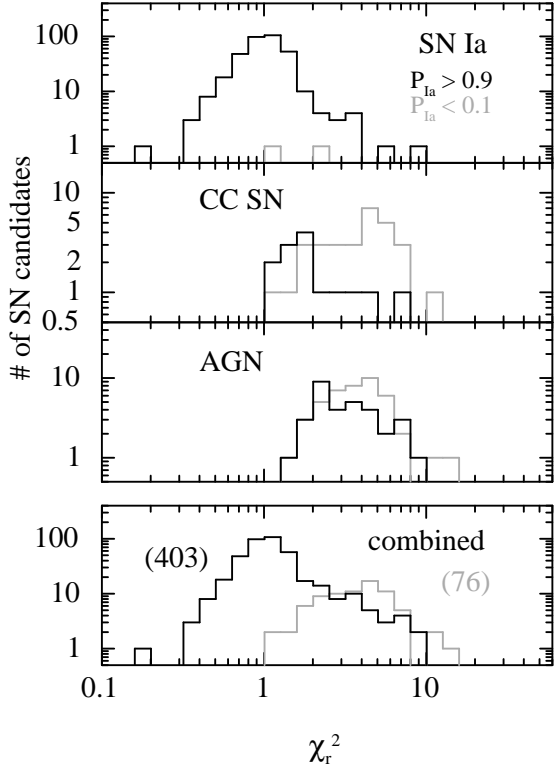}
  \end{center}
  \caption{Histograms of best-fit \chir\ values for a \snia\ model for
    \pia~$\geq 0.9$ (black) and \pia~$\leq 0.1$ (gray) for the
    spectroscopically confirmed (from top to bottom) 1) \snia, 2) \ccsn, 3)
    \agn, and 4) all three samples combined.  Note that the vast majority of
    \snia\ have \pia~$\geq 0.9$.  The contaminating false-positives are the
    \ccsn\ and \agn\ represented by the black histograms with $P_{\mathrm{Ia}}
    \ge 0.9$, and there are only a small number of those sources in our
    sample.}
  \label{fig:chisq_hist}
\end{figure}

\begin{figure}[htb]
  \begin{center}
    \includegraphics[angle=-90, width=.45\textwidth]{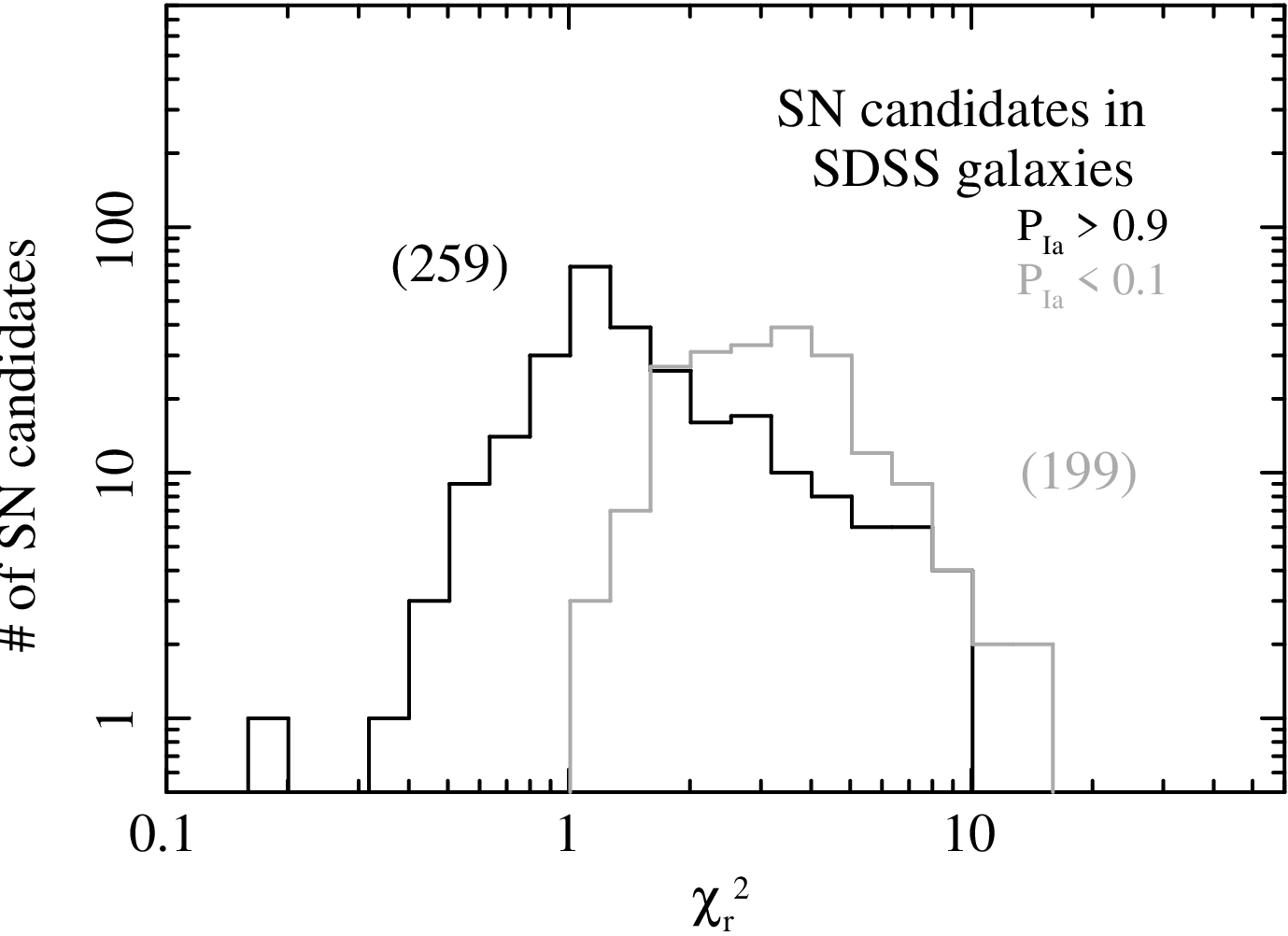}
  \end{center}
  \caption{Histograms of best-fit \chir\ values for a \snia\ model for
    $P_{\mathrm{Ia}} \ge 0.9$ (black) and $P_{\mathrm{Ia}} \le 0.1$ (gray) for
    the \sn\ candidates in \sdss\ spectroscopic galaxies.}
  \label{fig:chisq_hist_sdss}
\end{figure}

\begin{figure}[tb]
  \begin{center}
    \includegraphics[angle=-90, width=.45\textwidth]{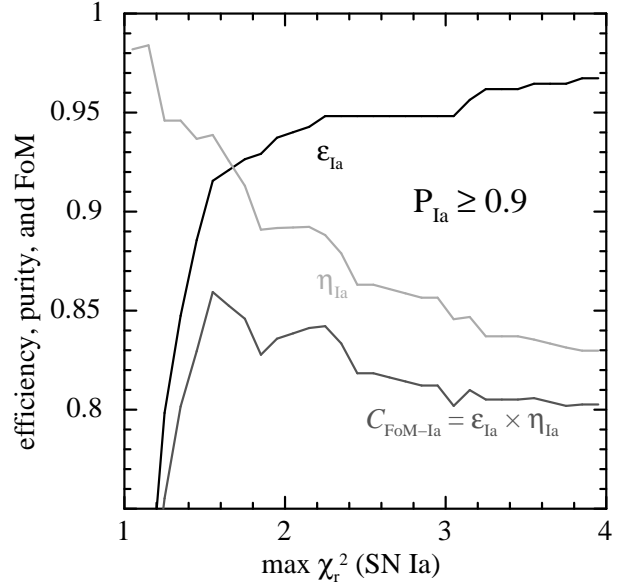}
  \end{center}
  \caption{The efficiency, purity, and figure of merit for the
    spectroscopically-confirmed \sn\ as functions of the maximum-allowed
    \chir\ for \pia~$\geq 0.9$.}
  \label{fig:eff_pur_fom}
\end{figure}

\begin{figure}[htb]
  \begin{center}
    \includegraphics[angle=-90, width=.45\textwidth]{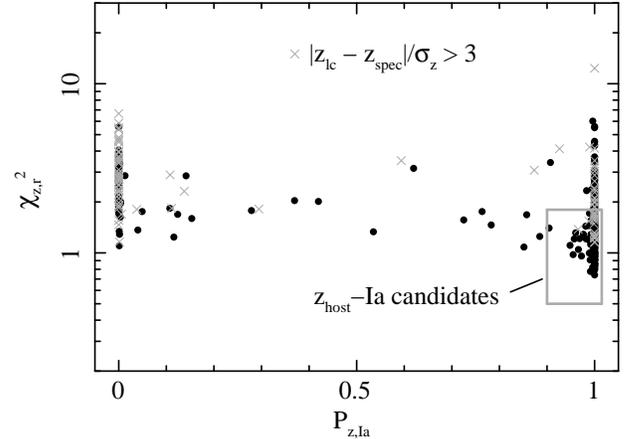}
  \end{center}
  \caption{\pzia\ vs \chizr\ for the 445 photometric candidates with galaxy
    spectroscopic redshifts.  Candidates that do not meet the light curve
    \photoz\ cut ($|$\zlc\ -- \zspec$|$/$\sigma_z < 3$) are shown as crosses.
    The 210 \zhost -Ia candidates identified are bounded by the red box shown
    in the lower right.}
  \label{fig:photo_selection_+hostz}
\end{figure}

  We also show in Figure~\ref{fig:chisq_hist} histograms of the \chir\ values
  for the same sources for \pia~$\ge 0.9$.  Of the $\niacut = 367$
  spectroscopic \snia\ that pass our light curve quality cuts, 357 sources
  have \pia~$\ge 0.9$.  There are also 14 \ccsn\ and 32 \agn\ with \pia~$\ge
  0.9$.

  For estimating $\puria$, we apply the penalty factor only on the
  \ccsn\ sample where the bias is more significant.  Almost all of the
  spectroscopic \agn\ confirmation came from \sdss\ quasar spectroscopy
  \citep{richards02} and not from our own targeting, so we assume that this
  sample is unbiased.  The expression for the efficiency is given in
  Eq.\ref{equ:eff}.  We write the purity explicitly as,
\begin{equation}
  \label{equ:pur_spec}
  \puria = \frac{\niatrue}{\niatrue + \mathcal{W}_{\rm{Ia,CC}}^{\rm{false}}
    \mathcal{N}_{\rm{Ia,CC}}^{\rm{false}} + \mathcal{N}_{\rm{Ia,AGN}}^{\rm{false}}},
\end{equation}
  where we have assumed $\mathcal{W}_{\rm{Ia,AGN}}^{\rm{false}} = 1$.  The
  penalty factor $\mathcal{W}_{\rm{Ia,CC}}^{\rm{false}}$ can be estimated from
  the histograms shown in the bottom panel of Figure~\ref{fig:chisq_hist} and
  Figure~\ref{fig:chisq_hist_sdss}.  Specifically, we have
  $\mathcal{W}_{\rm{Ia,CC}}^{\rm{false}} = (403/76)/(259/199) = 4.1$ using the
  same method as for the case with the spectroscopic redshift prior.  We show
  in Figure~\ref{fig:eff_pur_fom} the efficiency and purity as a function of
  the maximum-allowed \chir\ value.  Also shown is the figure of merit, which
  exhibits a broad maximum at $\fomia = 0.86$.  At \chir~$\sim 1.6$, the
  efficiency and purity are $\sim 92$\% and $\sim 94$\%, respectively.

\section{SDSS-II Photometric SN~Ia Candidates}
\label{sec:photo-ia}

\begin{deluxetable*}{rrrrrrrr}
  \tabletypesize{\footnotesize}
  \tablewidth{0pt}
  \tablecaption{SDSS-II \zhost -Ia Candidates\tablenotemark{a}\label{tbl:zhostia}}
  \tablehead{
    \colhead{SDSS ID\tablenotemark{b}} &
    \colhead{RA\tablenotemark{c}} &
    \colhead{Dec\tablenotemark{c}} &
    \colhead{\zspec} &
    \colhead{\av} &
    \colhead{\dmB} &
    \colhead{\pzia} &
    \colhead{\chizr}
  }
  \startdata
   703  & $-23.782080$ & $ +0.650725$  & $ 0.3000 \pm  0.0100$ & $ 0.04^{+0.16}_{-0.18}$ & $ 0.70^{+0.13}_{-0.08}$ &  1.000  &  0.99  \\
   779  & $ 26.673738$ & $ -1.020580$  & $ 0.2377 \pm  0.0005$ & $ 0.21^{+0.13}_{-0.13}$ & $ 0.85^{+0.10}_{-0.09}$ &  1.000  &  0.80  \\
   841  & $ 48.495991$ & $ -1.010015$  & $ 0.2991 \pm  0.0005$ & $-0.17^{+0.16}_{-0.18}$ & $ 1.02^{+0.20}_{-0.16}$ &  1.000  &  0.99  \\
  1415  & $  6.106480$ & $ +0.599307$  & $ 0.2119 \pm  0.0002$ & $ 0.66^{+0.13}_{-0.13}$ & $ 0.76^{+0.09}_{-0.08}$ &  1.000  &  0.93  \\
  1461  & $ 24.372675$ & $ +0.209735$  & $ 0.3407 \pm  0.0005$ & $ 0.33^{+0.11}_{-0.11}$ & $ 1.08^{+0.12}_{-0.12}$ &  1.000  &  1.07  \\
  1595  & $-38.432114$ & $ -0.554060$  & $ 0.2136 \pm  0.0005$ & $ 0.07^{+0.09}_{-0.09}$ & $ 1.03^{+0.08}_{-0.08}$ &  1.000  &  1.56  \\
  1748  & $ -6.887835$ & $ -0.482495$  & $ 0.3397 \pm  0.0001$ & $ 0.52^{+0.20}_{-0.19}$ & $ 0.83^{+0.16}_{-0.13}$ &  0.996  &  1.00  \\
  1775  & $-41.006622$ & $ -1.009430$  & $ 0.3050 \pm  0.0100$ & $-0.27^{+0.17}_{-0.17}$ & $ 1.26^{+0.15}_{-0.15}$ &  1.000  &  1.07  \\
  1835  & $-47.335869$ & $ +1.071860$  & $ 0.2716 \pm  0.0100$ & $-0.19^{+0.19}_{-0.19}$ & $ 1.28^{+0.22}_{-0.20}$ &  1.000  &  1.31  \\
  \nodata \\
  \enddata
  \tablenotetext{a}{Full table is published in its entirety in the electronic
    edition of The Astrophysical Journal.  A portion is shown here for
    guidance regarding its form and content.}
  \tablenotetext{b}{Internal SN candidate designation.}
  \tablenotetext{c}{Coordinates are J2000.  Right ascension is given in
    decimal degrees defined in the range [$-180^\circ$, $+180^\circ$].}
\end{deluxetable*}

\begin{deluxetable*}{rrrrrrrr}
  \tabletypesize{\footnotesize}
  \tablewidth{0pt}
  \tablecaption{SDSS-II Photo-Ia Candidates\tablenotemark{a}\label{tbl:photoia}}
  \tablehead{
    \colhead{SDSS ID\tablenotemark{b}} &
    \colhead{RA\tablenotemark{c}} &
    \colhead{Dec\tablenotemark{c}} &
    \colhead{\zlc} &
    \colhead{\av} &
    \colhead{\dmB} &
    \colhead{\pia} &
    \colhead{\chir}
  }
  \startdata
   822  & $ 40.560776$ & $ -0.862157$  & $ 0.167^{+0.065}_{-0.050}$ & $ 0.51^{+0.40}_{-0.47}$ & $ 1.24^{+0.14}_{-0.14}$ &  1.000  &  1.38  \\
   859  & $ -9.448275$ & $ +0.386555$  & $ 0.305^{+0.025}_{-0.036}$ & $ 0.04^{+0.28}_{-0.31}$ & $ 0.77^{+0.13}_{-0.09}$ &  1.000  &  1.25  \\
   904  & $ 21.095400$ & $ -0.124883$  & $ 0.288^{+0.029}_{-0.026}$ & $-0.26^{+0.28}_{-0.34}$ & $ 1.10^{+0.22}_{-0.17}$ &  0.999  &  1.00  \\
  1158  & $ 17.275431$ & $ -0.352185$  & $ 0.694^{+0.006}_{-0.063}$ & $-0.58^{+0.50}_{-0.31}$ & $ 1.55^{+0.19}_{-0.29}$ &  1.000  &  1.01  \\
  1243  & $-18.340113$ & $ -0.764753$  & $ 0.188^{+0.100}_{-0.102}$ & $ 0.89^{+0.73}_{-0.67}$ & $ 0.69^{+0.09}_{-0.06}$ &  1.000  &  1.47  \\
  1285  & $-38.216843$ & $ +0.543195$  & $ 0.345^{+0.049}_{-0.078}$ & $ 0.21^{+0.47}_{-0.38}$ & $ 1.06^{+0.25}_{-0.20}$ &  1.000  &  1.01  \\
  1302  & $ 53.654808$ & $ +0.891903$  & $ 0.282^{+0.039}_{-0.042}$ & $-0.10^{+0.26}_{-0.34}$ & $ 0.82^{+0.08}_{-0.08}$ &  1.000  &  1.23  \\
  1342  & $-13.472480$ & $ +0.117010$  & $ 0.299^{+0.046}_{-0.050}$ & $ 0.03^{+0.28}_{-0.31}$ & $ 1.18^{+0.15}_{-0.14}$ &  1.000  &  0.90  \\
  1354  & $ -5.197145$ & $ +0.089970$  & $ 0.283^{+0.046}_{-0.056}$ & $ 0.30^{+0.45}_{-0.55}$ & $ 1.50^{+0.20}_{-0.27}$ &  0.999  &  0.91  \\
  \nodata \\
  \enddata
  \tablenotetext{a}{Full table is published in its entirety in the electronic
    edition of The Astrophysical Journal.  A portion is shown here for
    guidance regarding its form and content.}
  \tablenotetext{b}{Internal SN candidate designation.}
  \tablenotetext{c}{Coordinates are J2000.  Right ascension is given in
    decimal degrees defined in the range [$-180^\circ$, $+180^\circ$].}
\end{deluxetable*}

\begin{figure}[htb]
  \begin{center}
    \includegraphics[angle=-90, width=.45\textwidth]{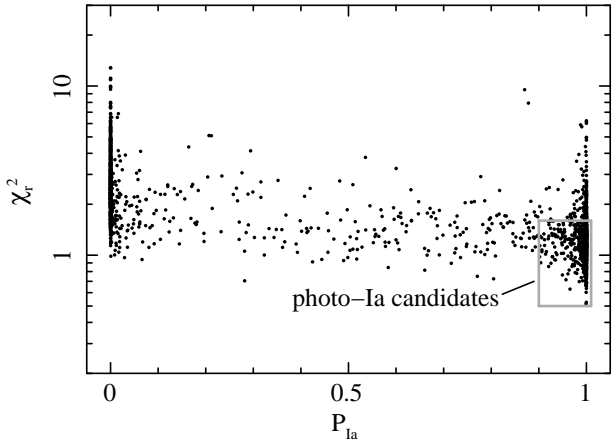}
  \end{center}
  \caption{\pia\ vs \chir\ for the 2776 photometric candidates with no
    spectroscopic information.  The 860 photometric \snia\ candidates are
    bounded by the gray box shown in the lower right.}
  \label{fig:photo_selection_nohostz}
\end{figure}

\begin{figure}[tb]
  \begin{center}
    \includegraphics[angle=0, width=.45\textwidth]{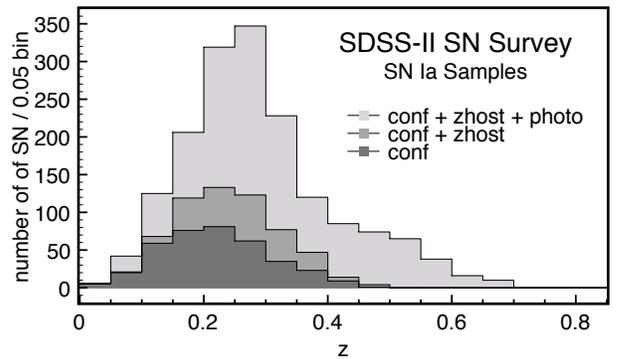}
  \end{center}
  \caption{Redshift distributions of the three \snia\ samples.}
  \label{fig:photo_zdist}
\end{figure}

  We now evaluate the light curves of the 445 candidates with spectroscopic
  redshift measurements of their host galaxies.  Their \sn\ types are unknown
  because there were no spectroscopic observations of these objects.
  Selection with \pzia~$\ge 0.90$, \chizr~$\le 1.8$, and $|$\zlc\ --
  \zspec$|$/$\sigma_z < 3$ results in 210 candidates shown in
  Figure~\ref{fig:photo_selection_+hostz}.  Based on the analysis presented in
  \S~\ref{subsec:withz}, we expect this sample to have an efficiency of $\sim
  89$\%, purity of $\sim 94\%$, and a figure-of-merit of $\sim 0.84$.  We
  refer to this sample of 210 candidates as the ``\zhost -Ia sample''.  Their
  candidate {\small ID}, coordinates, spectroscopic redshifts, and light curve
  fit results are listed in Table~\ref{tbl:zhostia}.

  From the 2776 candidates with no spectroscopy, identifying sources with
  \pia~$\ge 0.90$ and \chir~$\le 1.6$ results in $860$ purely-photometric
  \snia\ candidates, which we refer to as the ``photo-Ia sample''.  The
  selection is shown in Figure~\ref{fig:photo_selection_nohostz}.  We expect
  this sample to have an efficiency of $\sim 92\%$, a purity of $\sim 94\%$,
  and a figure-of-merit of $0.86$.  Its redshift distribution is shown in
  Figure~\ref{fig:photo_zdist}.  The mean redshift of the photo-Ia sample is
  $\bar{z} = 0.31$ compared to $\bar{z} = 0.22$ for the spectroscopically
  comfirmed sample.  The full list of candidates is provided in
  Table~\ref{tbl:photoia}.  In addition to their coordinates, we provide the
  photometric light curve redshifts \zlc\ marginalized over all the other
  parameters.  The reliability of these values is discussed in the following
  section.

  The light curves of these candidates, as well as all of the other
  \sn\ candidates, will be made available soon as part of the
  \sdssii\ \sn\ Survey Data Release.

\section{Photometric Redshifts and Distances}
\label{sec:redshift}

\begin{figure}[tb]
  \begin{center}
    \includegraphics[angle=-90, width=.40\textwidth]{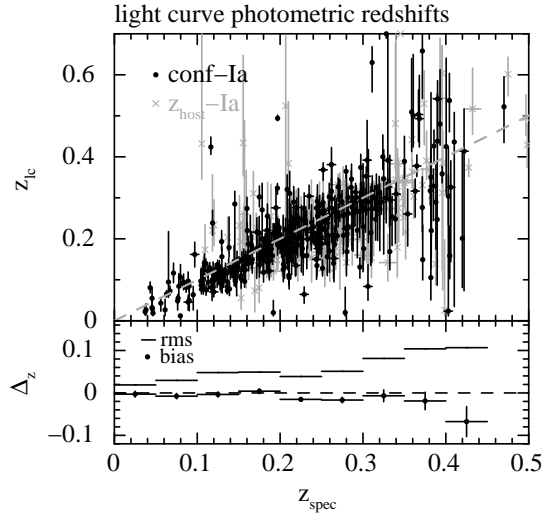}
  \end{center}
  \caption{Comparisons of \zspec\ and \zlc\ with a flat redshift prior for the
    spectroscopic \snia\ sample.  The 387 \snia\ that pass the light curve
    quality cuts are shown in black while the 210 \zhost -Ia are indicated by
    gray crosses in the top panel.  The bottom panel shows the mean $\Delta_z
    = (z_{\rm{lc}} - z_{\rm{spec}})/(1 + z_{\rm{spec}})$ values in black
    circles and the \rms\ as horizontal bars in bins of 0.05 for the combined
    \snia\ + \zhost -Ia samples.  The magnitude of the bias $|\Delta_z|$ is
    less than $0.02$ for $z < 0.4$.}
  \label{fig:zspec_zlc-flat}
\end{figure}

\begin{figure}[tb]
  \begin{center}
    \includegraphics[angle=-90, width=.40\textwidth]{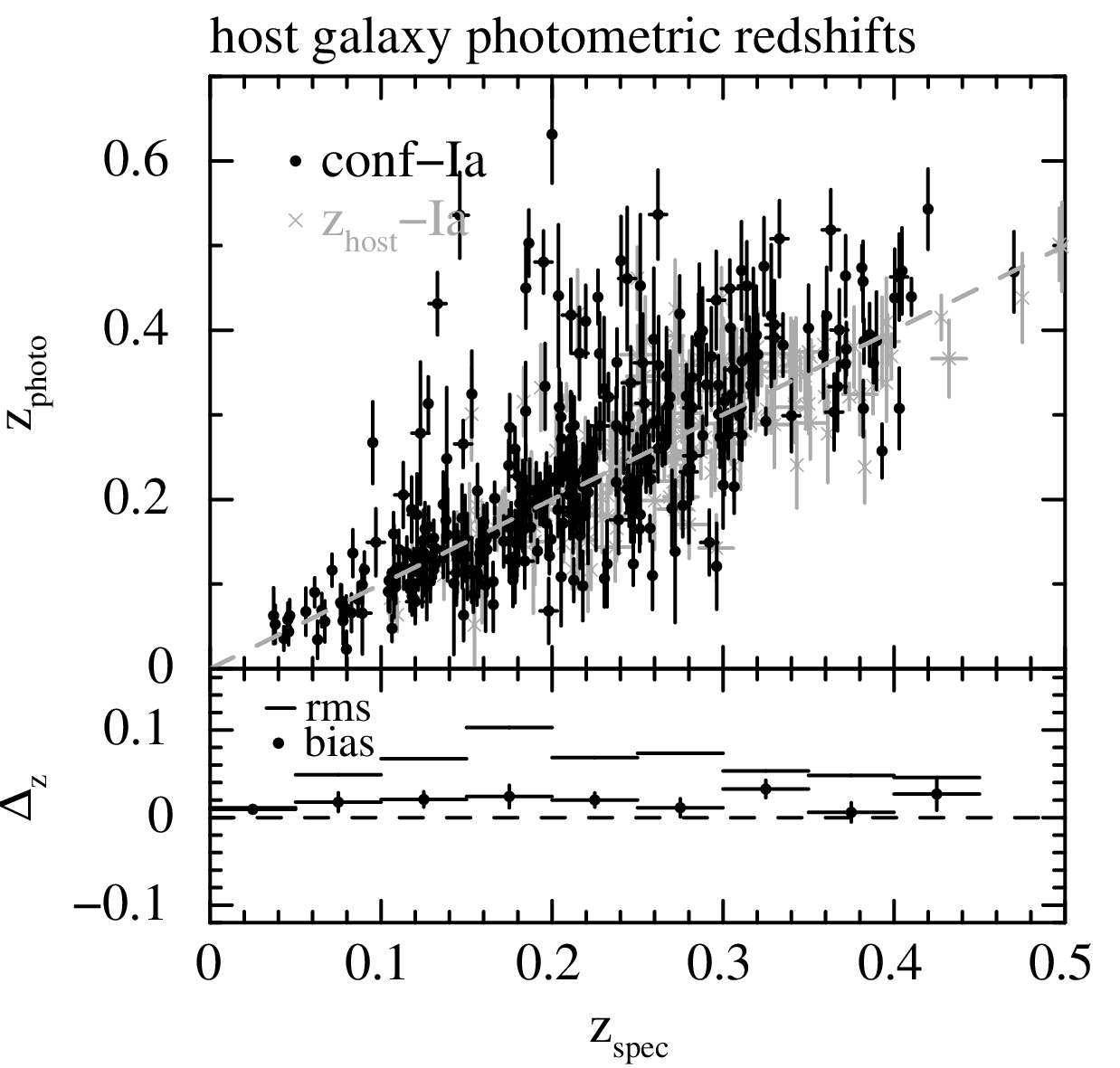}
  \end{center}
  \caption{Comparisons of \zspec\ and \zphoto, the photometric redshift of the
    \snia\ host galaxies from \citet{oyaizu08}.  There are fewer points here
    than in Figure~\ref{fig:zspec_zlc-flat} because there are many \snia\ with
    hosts that are below the detection limit of \sdss, and some galaxies are
    classfied as stars and therefore do not have \zphoto\ values.  The bottom
    panel shows the mean $\Delta_z$ values in black circles and the \rms\ as
    horizontal bars in bins of 0.05.}
  \label{fig:zspec_zphoto}
\end{figure}

\begin{figure*}[tb]
  \begin{center}
    \includegraphics[angle=-90, width=.70\textwidth]{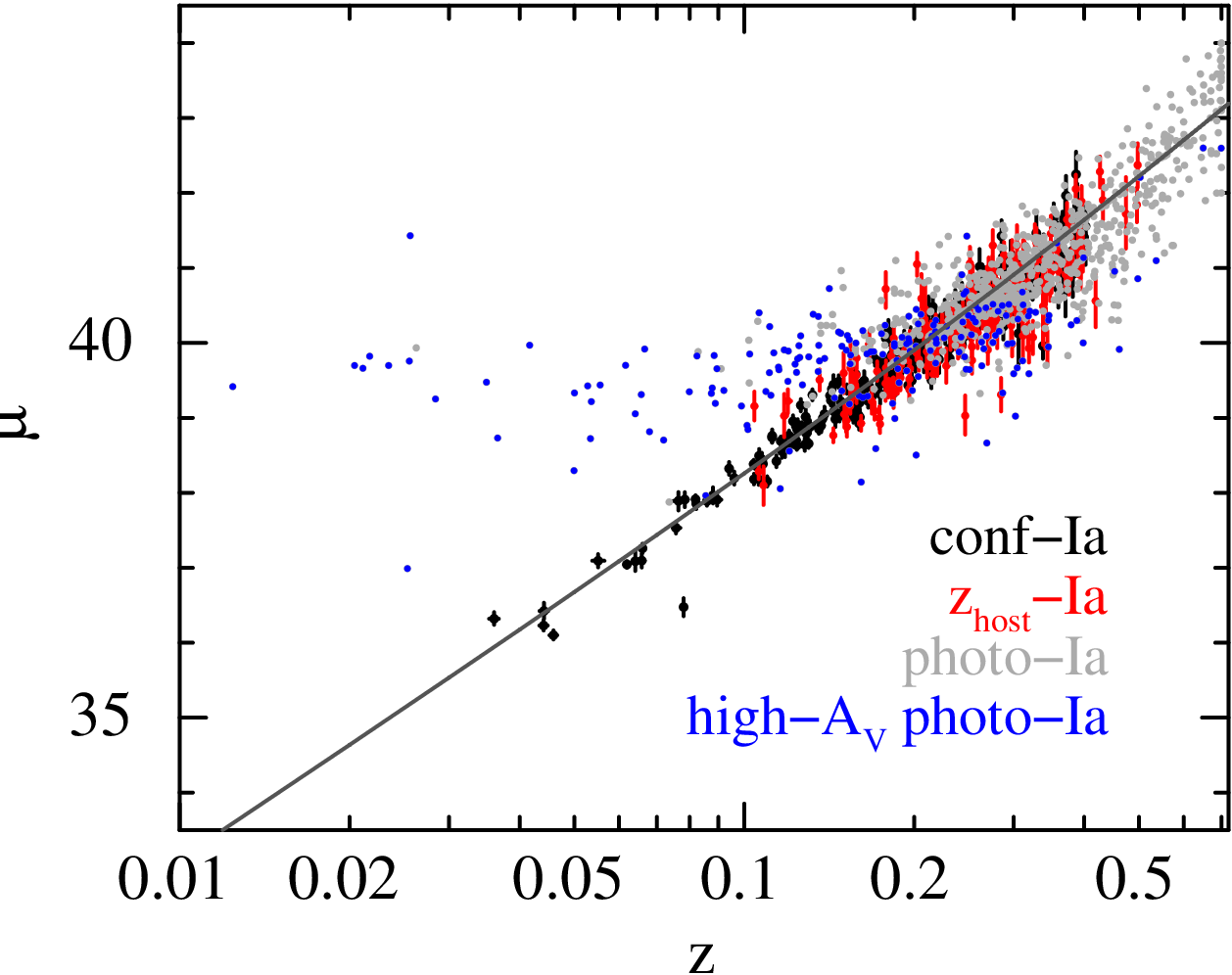}
  \end{center}
  \caption{Hubble diagram of the three \snia\ samples (conf-Ia in black,
    \zhost -Ia in red, and photo-Ia in light gray and blue in the online color
    version).  The dark gray line represents \lcdm.  Spectroscopic redshift
    priors are used for the conf-Ia and \zhost -Ia samples.  Flat redshifts
    priors are used for the photo-Ia sample.  The redshift and distance of the
    photo-Ia are significantly correlated and their uncertainties are not
    shown for clarity.  The outliers at low-$z$ are probably due to
    \ccsn\ that are mis-classified as high-\av\ ($A_V > 1$) \snia, which are
    shown in blue.  Note that the majority of these points are significantly
    away from the \lcdm\ Hubble relation.}
  \label{fig:snia_hubble}
\end{figure*}

  The light curve redshifts \zlc\ are determined by marginalizing over the
  other four model parameters; \av, \tmax, \dmB, and $\mu$.  For each
  \sn\ candidate, the posterior probability distribution function is
  constructed from the \mcmc\ output.  The redshifts listed in
  Table~\ref{tbl:photoia} correspond to the median \zlc\ and the $\pm 34.1$\%
  ($1\sigma$) upper and lower limits.

  We compare the spectroscopic redshifts \zspec\ with \zlc\ for the conf-Ia
  and \zhost -Ia samples and with the host galaxy photometric redshifts
  \zphoto\ from \citet{oyaizu08} available in the \sdss\ {\small DR}8
  database.  As shown in Figure~\ref{fig:zspec_zlc-flat}, \zlc\ and
  \zspec\ are in agreement with $|\Delta_z| < 0.02$ ($\Delta_z \equiv
  (z_{\rm{lc}} - z_{\rm{spec}})/(1 + z_{\rm{spec}})$) for $z_{\rm{spec}} <
  0.4$, but with a small redshift-dependent bias.  The \rms\ scatter is
  $\Delta_{z,\rm{RMS}} = 0.05$ below $z_{\rm{spec}} = 0.30$ and increases to
  $0.1$ at $z = 0.4$.

  The sign and magnitude to this bias is similar to those found by
  \citet{kessler10a}, who analyzed a subset of the higher
  \signoi\ \sdssii\ \snia\ light curves presented here using both \mlcs\ and
  \saltii.  Interestingly, a similar bias is seen in their simulations.
  \citet{rodney10a} do not quote a value for the bias, but they state that a
  line with a slope of unity fits the \zspec\ vs.\ \zlc\ values for the
  first-year \sdssii\ \snia\ sample with a \chir $= 0.98$.

  We also show in Figure~\ref{fig:zspec_zphoto} a comparison of \zspec\ with
  the host galaxy photometric redshift \zphoto\ from \citet{oyaizu08}.  Here,
  there is a nearly constant bias of $\Delta_z \sim 0.03$ with an
  \rms\ scatter of $\Delta_{z,\rm{RMS}} \sim 0.05 - 0.10$.

\begin{figure}[tb]
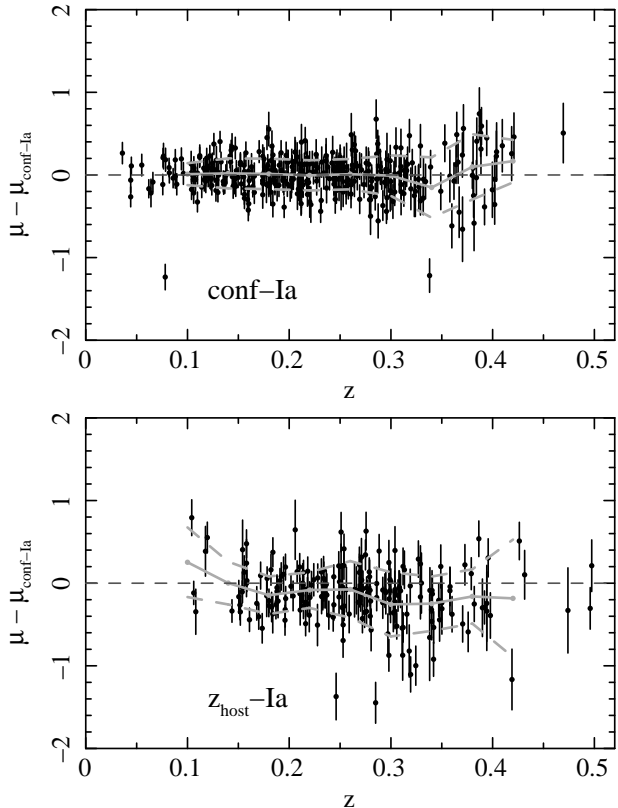

  \begin{center}
    \includegraphics[angle=-90, width=.45\textwidth]{fig21a.ps}
    \includegraphics[angle=-90, width=.45\textwidth]{fig21b.ps}
  \end{center}
  \caption{(Top) The Hubble residuals of the conf-Ia sample relative to a
    simple quadratic Hubble relation.  The solid line represents the mean
    residual and the dashed lines represent upper and lower rms values
    relative to the mean.  The scatter ranges from $\sigma_\mu \sim 0.13$~mag
    to $\sigma_\mu \sim 0.30$~mag in the redshift interval $0.1 < z < 0.4$.
    (Bottom) Same except for the \zhost-Ia sample.  The same quadratic has
    been subtracted from the measured distance modulus.  The scatter here is
    larger and ranges from $\sigma_\mu \sim 0.2$~mag to $\sigma_\mu \sim
    0.4$~mag in the same redshift range.  There is also a small
    redshift-dependent offset.}
  \label{fig:snia_hubble_res}
\end{figure}

\begin{figure}[tb]
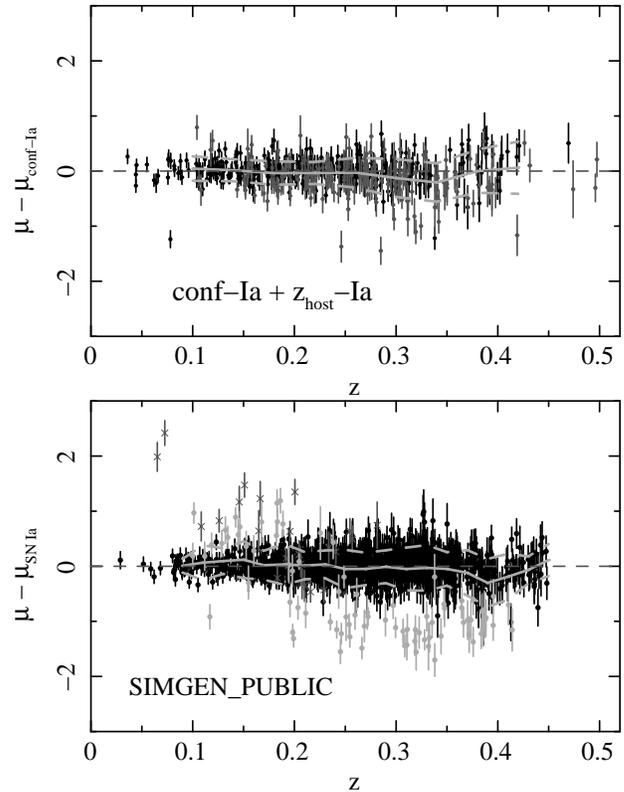

  \begin{center}
    \includegraphics[angle=-90, width=.45\textwidth]{fig22a.ps}
    \includegraphics[angle=-90, width=.45\textwidth]{fig22b.ps}
  \end{center}
  \caption{(Top) Same as in Figure~\ref{fig:snia_hubble_res} for the combined
    conf-Ia + \zhost-Ia sample, which are labeled in black and light gray,
    respectively.  The same quadratic function $\mu_{\rm{conf-Ia}}(z)$ has
    been subtracted from the measured distance modulus.  The rms scatter is
    slightly larger than that of the conf-Ia sample only.  (Bottom) Simulated
    \sdssii\ \sn\ from K10b.  The black, light gray, and dark gray points
    represent \snia, \snii, and \snibc, respectively, which pass all of the
    photometric \snia\ cuts (\pia~$\ge 0.9$ and \chizr~$<1.0$).  The residuals
    shown are relative to a quadratic fit to the simulated \snia\ sample only,
    whereas the rms scatter shown is for the full sample.  Note the slight
    redshift-dependent bias relative to the \snia\ mean.}
  \label{fig:snia_hubble2_res}
\end{figure}

\begin{figure}[tb]
  \begin{center}
    \includegraphics[angle=-90, width=.45\textwidth]{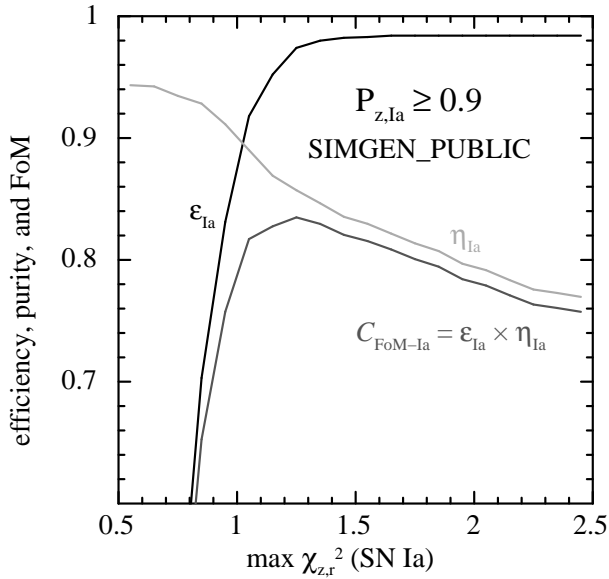}
  \end{center}
  \caption{Same as in Figure~\ref{fig:zeff_pur_fom} for simulated
    \sdssii\ \sn\ from K10b except with an additional constraint of
    \signoi~$>7$ in two bands to approximately match the \zhost -Ia sample.}
  \label{fig:simgen_fom}
\end{figure}

  We show in Figure~\ref{fig:snia_hubble} the Hubble diagram of the 350
  conf-Ia, 210 \zhost -Ia, and 860 photo-Ia samples.  Distance modulus
  residuals of the conf-Ia and \zhost -Ia samples relative to a simple
  quadratic fit are shown in Figure~\ref{fig:snia_hubble_res}.  For the
  conf-Ia sample, the scatter around the mean Hubble relation is $\sigma_\mu =
  0.13$~mag at $z = 0.1$ and increases monotonically to $\sigma_\mu =
  0.30$~mag at $z = 0.4$.  The same Hubble relation was subtracted from the
  \zhost-Ia sample, which is shown in the right panel of
  Figure~\ref{fig:snia_hubble_res}.  There is a noticeably larger scatter with
  $\sigma_\mu = 0.2 - 0.4$~mag in the same redshift range.  This is most
  likely due to contamination from non-Ia events, which we have estimated to
  be at the level of $\sim 6$\% (approximately 1 out of 16 events in this
  sample is likely to be a \ccsn).  The slight deviation of the mean from zero
  is not statistically significant.

  The Hubble diagram of the photo-Ia sample shows extreme outliers below $z
  \sim 0.1$.  All of these points are significantly above the \lcdm\ Hubble
  relation, and are most likely \ccsn\ that are mis-classified as \snia.  In
  fact, the majority of these events are classified by \psnid\ as
  extremely-underluminous, high-extinction ($A_V \ga 1$) \snia.  Since the
  underlying extinction distribution of \snia\ follows the relation $\propto
  e^{-A_V/\tau_V}$ with $\tau_V \sim 0.33$ \citep{kessler09a}, and given the
  smaller number of confirmed \snia\ in the same redshift interval, it is
  unlikely that all of these outlier events are underluminous, high-extinction
  \snia.  Selecting only the candidates with $A_V < 1$ eliminates most of
  these outliers at the cost of a somewhat reduced efficiency, but
  measurements of their host galaxy redshifts will also significantly help
  distinguish their types.

  At higher redshifts, the mean Hubble relation of the photo-Ia sample is
  consistent with the conf-Ia and \zhost -Ia samples, but with a significantly
  larger scatter.  Above $z \sim 0.2$, the rms scatter is $\sigma_\mu \sim 0.5
  - 0.7$~mag, which is about a factor of $\sim 2$ larger than the scatter in
  the conf-Ia and \zhost -Ia samples in the same redshift range.

\section{Comparisons with Simulations}
\label{sec:sims}

  The Hubble diagram for the combined conf-Ia + \zhost-Ia sample is shown in
  the top panel of Figure~\ref{fig:snia_hubble2_res}.  The scatter is
  $\sigma_\mu = 0.2$~mag at $z = 0.1$ and increases to $\sigma_\mu = 0.4$~mag
  at $z = 0.4$, which is slightly larger than the scatter of the conf-Ia
  sample.

  This degradation is probably due to contamination by \ccsn\ events, but to
  test this hypothesis, we analyzed the sample of simulated \sdssii\ \sn\ from
  K10b.  This simulation corresponds to 10 three-season search campaigns, and
  uses the actual seeing, photometric zeropoints, and weather from our
  observing seasons.  The right panel in Figure~\ref{fig:snia_hubble2_res}
  shows the Hubble diagram using all events that pass the same light curve
  quality cuts, as well as identical selection criteria in \pzia -
  \chizr\ space.  Specifically, we select \snia\ candidates using \pzia~$\ge
  0.9$ and \chizr~$< 1.0$, which is approximately where the efficiency and
  purity are equal at $\sim 0.90$ for this simulation.  The efficiency,
  purity, and figure-of-merit curves are shown in Figure~\ref{fig:simgen_fom}.
  The average \signoi\ of the \zhost -Ia sample is higher than that of the
  simulations, so we require in the simulations \signoi~$>7$ in at least two
  of the $gri$ bands.  The purity of 90\% for this selection is slightly lower
  than the estimated purity of the \zhost -Ia sample.

  The \snia\ Hubble digram was fitted to a quadratic function and the Hubble
  residuals of all candidates classified as \snia\ are shown in the bottom
  panel of Figure~\ref{fig:snia_hubble2_res}.  Here, the \ccsn\ events are
  shown in dark (\snibc) and light gray (\snii) points.  These false-positives
  are adding scatter and a small redshift-dependent systematic shift relative
  to the \snia\ distances, which are represented by black points.  The Hubble
  scatter around the mean for this simulation is $\sigma_\mu = 0.2 - 0.4$~mag,
  which is similar to the that of the \zhost-Ia sample over the entire
  redshift range.  The larger scatter seen in the conf-Ia + \zhost -Ia sample
  is, therefore, most likely due to mis-classified \ccsn\ as reproduced in
  these simulations.

  This set of simulated \sdssii\ \sn\ also includes a spectroscopic
  \snia\ sample selected based on our spectroscopic follow-up strategies, and
  represents our conf-Ia.  The Hubble residual scatter of this spectroscopic
  sample ranges from $\sigma_\mu \sim 0.13$~mag to $\sigma_\mu \sim 0.30$~mag
  in the redshift interval $0.1 < z < 0.4$, which is nearly identical to the
  observed scatter of the conf-Ia sample.

\section{Summary and Conclusions}
\label{sec:summary}

  We have identified $1070$ photometric \snia\ candidates from the
  \sdssii\ \sn\ Survey data.  This sample is more than three times larger than
  the spectroscopically confirmed \snia\ sample with good light curves, and is
  estimated to include $\sim 91$\% of all \snia\ candidates detected by the
  survey with a purity of $\sim 94$\% ($\sim 6$\% contamination).  This
  estimate of the purity, however, is based on a limited number of
  spectroscopically confimred \ccsn, most of which are nearby, bright events
  and are therefore not representative of the majority of the contaminating
  events.  As shown in Figure~\ref{fig:snr}, the majority of our photometric
  candidates have peak $r$-band \signoi $< 10$, where we have only a handful
  of spectroscopic \sn\ candidates.  To obtain a better characterization of
  the contaminating sources, confirmation is needed for a much larger sample
  of faint \ccsn\ that are comparable in apparent brightness to the photo-Ia
  sample.  As also advocated by \citet{richards11}, future surveys that rely
  on photometric identification should obtain spectra of \sn\ candidates over
  the full range of the \signoi\ of the photometric candidates of interest.

  The Hubble digram with photometric classification and host galaxy
  spectroscopic redshift priors show a slight increase in scatter over the
  confirmed \snia\ sample, which is consistent with them being due to
  mis-classified \ccsn.  There is no significant redshift-dependent offset in
  the derived distances compared to the conf-Ia sample.  Simulations confirm
  these findings.

  Photometric redshifts estimated from the multi-band light curves are
  unbiased below $z \sim 0.2$ with an rms dispersion of $\sigma_z \sim 0.05$.
  There is a redshift-dependent bias above $z \sim 0.2$ where the mean
  redshift difference $\langle z_{\mathrm{lc}} - z_{\mathrm{photo}} \rangle$
  is between $-0.04$ and $-0.02$.  The rms dispersion is $\sigma_z \sim 0.05 -
  0.10$.  The Hubble diagram of the photo-Ia sample also exhibits outliers and
  redshift-dependent biases.  Although the distance and redshift accuracies at
  present are not yet sufficient for cosmology, the large sample can still be
  used for studies of the \snia\ rate as a function of redshift, correlations
  between \sn\ light curves and host galaxy properties, and other studies that
  do not involve joint constaints on both redshift and distance.

  We conclude that cosmology with future large-scale \sn\ surveys should at
  the minimum measure host galaxy spectroscopic redshifts for the Hubble
  digram.  A subset of the \sn\ candidates must be observed spectroscopically
  to study the photometric classification efficiency and purity.  Spectroscopy
  should target candidates with \signoi\ down to the magnitude limit where
  photometric classification is expected to work.  Cosmology with photometry
  alone, however, requires further investigation with realistic simulations in
  order to understand and characterize their systematic biases and
  uncertainties, and how they depend on the \snia\ candidate selection
  criteria.


  \acknowledgements We thank the anonymous referee, who has helped improve the
  presentation of the paper.  Funding for the \sdss\ and \sdssii\ has been
  provided by the Alfred P. Sloan Foundation, the Participating Institutions,
  the National Science Foundation, the U.S. Department of Energy, the National
  Aeronautics and Space Administration, the Japanese Monbukagakusho, the Max
  Planck Society, and the Higher Education Funding Council for England. The
  \sdss\ Web Site is \verb9http://www.sdss.org/9.

  The \sdss\ is managed by the Astrophysical Research Consortium for the
  Participating Institutions. The Participating Institutions are the American
  Museum of Natural History, Astrophysical Institute Potsdam, University of
  Basel, Cambridge University, Case Western Reserve University, University of
  Chicago, Drexel University, Fermilab, the Institute for Advanced Study, the
  Japan Participation Group, Johns Hopkins University, the Joint Institute for
  Nuclear Astrophysics, the Kavli Institute for Particle Astrophysics and
  Cosmology, the Korean Scientist Group, the Chinese Academy of Sciences
  ({\small LAMOST}), Los Alamos National Laboratory, the Max-Planck-Institute
  for Astronomy ({\small MPIA}), the Max-Planck-Institute for Astrophysics
  ({\small MPA}), New Mexico State University, Ohio State University,
  University of Pittsburgh, University of Portsmouth, Princeton University,
  the United States Naval Observatory, and the University of Washington.

  The Hobby-Eberly Telescope (\het) is a joint project of the University of
  Texas at Austin, the Pennsylvania State University, Stanford University,
  Ludwig-Maximillians-Universit\"at M\"unchen, and Georg-August-Universit\"at
  G\"ottingen.  The \het\ is named in honor of its principal benefactors,
  William P. Hobby and Robert E. Eberly.  The Marcario Low-Resolution
  Spectrograph is named for Mike Marcario of High Lonesome Optics, who
  fabricated several optics for the instrument but died before its completion;
  it is a joint project of the Hobby-Eberly Telescope partnership and the
  Instituto de Astronom\'{\i}a de la Universidad Nacional Aut\'onoma de
  M\'exico.  The Apache Point Observatory 3.5-meter telescope is owned and
  operated by the Astrophysical Research Consortium.  We thank the observatory
  director, Suzanne Hawley, and site manager, Bruce Gillespie, for their
  support of this project.  The Subaru Telescope is operated by the National
  Astronomical Observatory of Japan.  The William Herschel Telescope is
  operated by the Isaac Newton Group, and the Nordic Optical Telescope is
  operated jointly by Denmark, Finland, Iceland, Norway, and Sweden, both on
  the island of La Palma in the Spanish Observatorio del Roque de los
  Muchachos of the Instituto de Astrofisica de Canarias.  Observations at the
  {\small ESO} New Technology Telescope at La Silla Observatory were made
  under programme {\small ID}s 77.A-0437, 78.A-0325, and 79.A-0715.  Kitt Peak
  National Observatory, National Optical Astronomy Observatory, is operated by
  the Association of Universities for Research in Astronomy, Inc. ({\small
    AURA}) under cooperative agreement with the National Science Foundation.
  The {\small WIYN} Observatory is a joint facility of the University of
  Wisconsin-Madison, Indiana University, Yale University, and the National
  Optical Astronomy Observatories.  The W.M.\ Keck Observatory is operated as
  a scientific partnership among the California Institute of Technology, the
  University of California, and the National Aeronautics and Space
  Administration.  The Observatory was made possible by the generous financial
  support of the W.M.\ Keck Foundation.  The South African Large Telescope of
  the South African Astronomical Observatory is operated by a partnership
  between the National Research Foundation of South Africa, Nicolaus
  Copernicus Astronomical Center of the Polish Academy of Sciences, the
  Hobby-Eberly Telescope Board, Rutgers University, Georg-August-Universit\"at
  G\"ottingen, University of Wisconsin-Madison, University of Canterbury,
  University of North Carolina-Chapel Hill, Dartmough College, Carnegie Mellon
  University, and the United Kingdom \salt\ consortium.  The Telescopio
  Nazionale Galileo ({\small TNG}) is operated by the Fundaci\'on Galileo
  Galilei of the Italian {\small INAF} {Istituo Nazionale di Astrofisica) on
    the island of La Palma in the Spanish Observatorio del Roque de los
    Muchachos of the Instituto de Astrof\'{\i}sica de Canarias.

\end{document}